\documentclass[fleqn,10pt]{wlscirep}

\usepackage[utf8]{inputenc}
\usepackage[T1]{fontenc}
\usepackage{times}
\usepackage{epsfig}
\usepackage{graphicx}
\usepackage{amsmath}
\usepackage{amssymb}
\usepackage{soul}
\usepackage{xcolor}

\usepackage{mathbbol}
\usepackage{multicol}
\usepackage[numbers,sort&compress]{natbib}
\usepackage{lineno}

\newcommand{\NCMRGBR}{31,602\,}
\newcommand{\NCALLS}{801,526\,} 
\newcommand{\NIMP}{9,472,708 } 
\newcommand{\NRUNS}{36 }
\newcommand{\NZVAR}{384 }
\newcommand{\NZVARPRUNED}{324 }
\newcommand{\RMSETHR}{1 mm }
\newcommand{\PGW}{p_{\text{GW}}}
\newcommand{\PSW}{p_{\text{SW}}}
\newcommand{\PGWVAL}{5\times10^{-8}}
\newcommand{\PSWVAL}{1.5\times10^{-10}}

\newcommand{\ACCESSIONNUMBER}{11350}
\title{Unsupervised ensemble-based phenotyping helps enhance the discoverability of genes related to heart morphology}

\author[1,2]{Rodrigo Bonazzola}
\author[3]{Enzo Ferrante}
\author[1,2]{Nishant Ravikumar}
\author[1,2]{Yan Xia}
\author[6,7]{Bernard Keavney}
\author[2]{Sven Plein}
\author[4]{Tanveer Syeda-Mahmood}
\author[1,2,5,*]{Alejandro F Frangi}
\affil[1]{Centre for Computational Imaging and Simulation Technologies in Biomedicine (CISTIB), School of Computing and School of Medicine, University of Leeds, Leeds, UK}
\affil[2]{Leeds Institute of Cardiovascular and Metabolic Medicine, School of Medicine, University of Leeds, Leeds, UK}
\affil[3]{Research Institute for Signals, Systems and Computational Intelligence, sinc(i), FICH-UNL / CONICET, Santa Fe, Argentina}
\affil[4]{IBM Almaden Research Center, San Jose, USA}
\affil[5]{Medical Imaging Research Center (MIRC), University Hospital Gasthuisberg. Cardiovascular Sciences and Electrical Engineering Departments, KU Leuven, Leuven, Belgium}
\affil[6]{Division of Cardiovascular Sciences, Faculty of Biology, Medicine and Health, University of Manchester, Manchester, UK}
\affil[7]{Manchester University NHS Foundation Trust, Manchester Academic Health Science Centre, Manchester, UK}

\affil[*]{a.frangi@leeds.ac.uk}

\begin{abstract}
Recent genome-wide association studies (GWAS) have been successful in identifying associations between genetic variants and simple cardiac parameters derived from cardiac magnetic resonance (CMR) images. However, the emergence of big databases including genetic data linked to CMR, facilitates investigation of more nuanced patterns of shape variability. Here, we propose a new framework for gene discovery entitled Unsupervised Phenotype Ensembles (UPE). UPE builds a redundant yet highly expressive representation by pooling a set of phenotypes learned in an unsupervised manner, using deep learning models trained with different hyperparameters. These phenotypes are then analyzed via (GWAS), retaining only highly confident and stable associations across the ensemble. We apply our approach to the UK Biobank database to extract left-ventricular (LV) geometric features from image-derived three-dimensional meshes. We demonstrate that our approach greatly improves the discoverability of genes influencing LV shape, identifying 11 loci with study-wide significance and 8 with suggestive significance. We argue that our approach would enable more extensive discovery of gene associations with image-derived phenotypes for other organs or image modalities.
\end{abstract}

\begin{document}

\flushbottom
\maketitle

\thispagestyle{empty}

\section*{Introduction}
Genome-wide association studies (GWAS) have remarkably accelerated discoveries of associations between genomic and complex traits ~\citep{ref_gwas_review}. In general, they analyse genetic variants (i.e the genotype) in a sample of individuals, to test their possible association with the presence of disease or with systematic changes in measurable traits, known broadly as phenotypes in this context. GWAS have already successfully identified genetic variants associated with a broad range of diseases and other complex traits, such as metabolic, anthropometric or behavioural ones. These findings have improved our understanding of the pathogenesis of disease, facilitating the development of better treatments, supporting drug discovery and assisting advances towards precision medicine.

Large-scale epidemiological imaging studies have correlated image-derived phenotypes (IDPs) to genetic data for identifying the genetic basis of organ structure and function in health and disease.
In cardiology, GWAS have been performed on clinically relevant quantitative indices of the left ventricle (LV), such as LV volumes, LV mass and LV ejection fraction,
as the diagnosis of patients with heart disease typically starts from quantitative analysis of this cardiac chamber ~\citep{ref_nayaung, ref_biffi}. Although there are discrepancies in the number of genetic loci associated with changes in LV IDPs from recently reported GWAS \citep{ref_nayaung, ref_pirruccello}, some consistent genetic factors have been established. 

These cardiac imaging genetics studies were based on traditional approaches, where handcrafted features characterising LV IDPs were first determined, before running GWAS to find the associated genetic loci. Although these IDPs have been clinically used to diagnose heart disease, they do not provide detailed representations of the chamber's morphology and its variation across the population. In this paper, we advance the view that shape features encoded in a learned latent space can provide a more refined imaging phenotype which turns out to be more informative than traditional measurements. When associated with genetic variation, this can provide novel insights into the genetic basis of cardiac structure and function.

\begin{figure*}
\includegraphics[width=\textwidth]{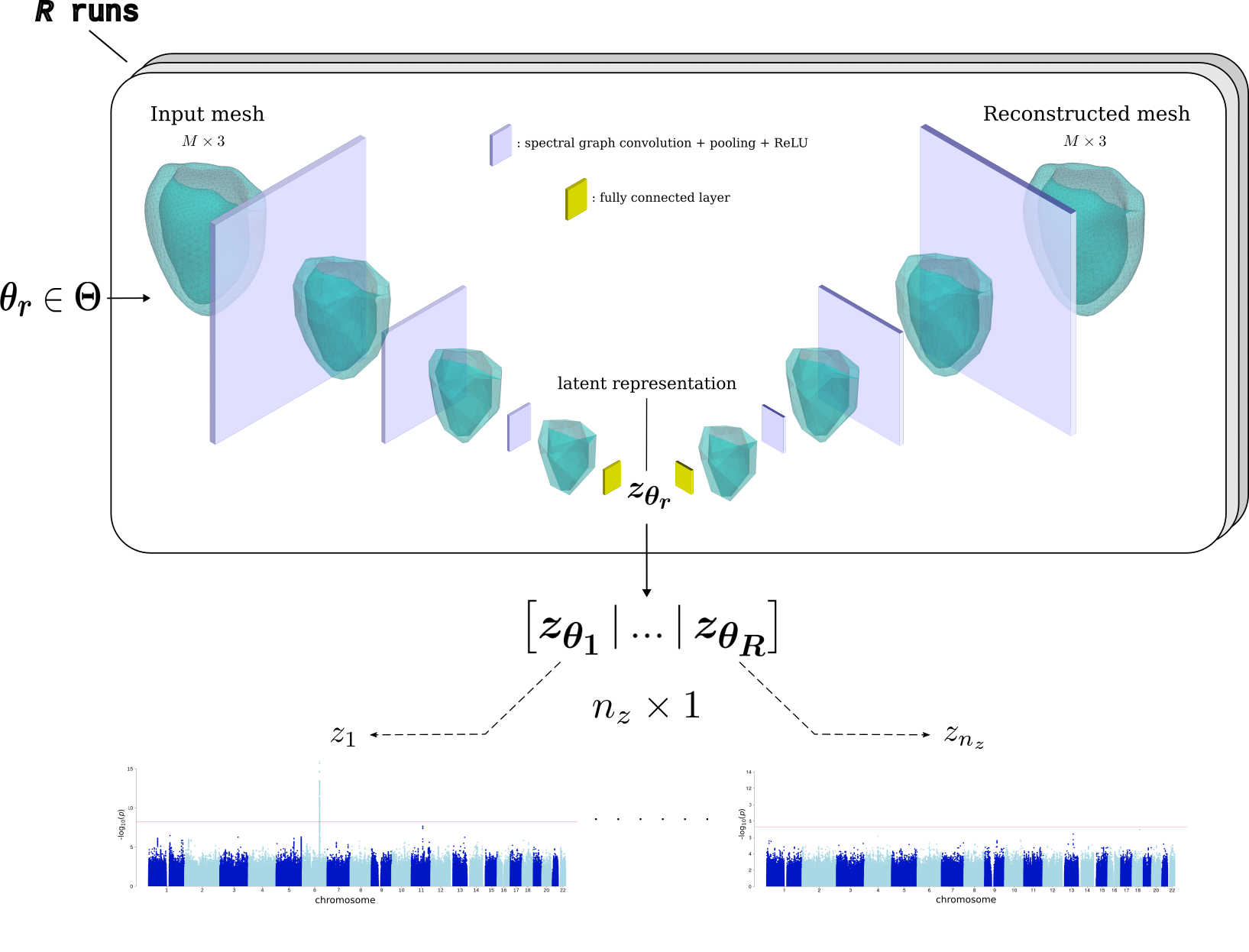}
\caption{Flowchart of the proposed Unsupervised Phenotype Ensembles (UPE) framework. First, a graph-convolutional autoencoder is trained and applied to our set of left-ventricular meshes (number of vertices $M=5220$) to produce low-dimensional representations of these shapes. In each layer, a representation with fewer vertices is obtained. 
The bottleneck $\textbf{z}_{\theta_{r}}$ of the autoencoder with hyperparameters $\theta_{r}$ is a $n_{z}^{r}$-dimensional vector for each run $r$ ($n_{z}^{r}\in\{8, 16\}$). The different latent vectors obtained for each run, $\{\textbf{z}_{\theta_{r}}\}_{r=1}^{R}$ are then tested in a GWAS component by component, for association with genetic variants.}
\label{fig:flowchart}
\end{figure*}

The unprecedented amount of linked genetic and cardiac imaging data available within the UK Biobank (UKB) \citep{ref_ukbb} facilitates the use of unsupervised machine learning techniques to automatically learn a set of features that best describe the morphology of the heart. The idea of using unsupervised learning to produce phenotypes that can be tested in GWAS has been previously explored by us in a conference paper \cite{ref_bonazzola2021miccai}, and also recently utilised in \citep{ref_kirchler2022transfergwas} in order to discover genetic loci related to phenotypes derived from retinal fundus images.

Atlas-based methods have been proposed to generate three-dimensional (3D) meshes representing cardiac anatomy from volumetric images \citep{ref_rahman, ref_zhuang_regis_2O10}. We build on top of these works, leveraging the latest advances in graph-convolutional neural networks \citep{ref_bronstein_geom_DL} to learn low-dimensional representations that consider mesh topology. While standard convolutional neural networks operate on domains with an underlying Euclidean or grid-like structure (e.g. images), geometric deep learning generalises convolutions to non-Euclidean domains such as graphs, meshes and manifolds, taking into account their topology and irregular structure. Previous studies employed mesh autoencoders in modelling the expression space of human face surfaces \citep{ref_coma}. Here we show that such models can enable anatomical variation in cardiac structures to be learned and correlated to genetic data.

In this work, we learn compact and non-linear representations of cardiac anatomy in an unsupervised setting via convolutional mesh autoencoders (CoMA). We hypothesise that the learned features can identify genetic loci that impact cardiac morphology due to their ability to explain shape variability across the population. We show that such representations can indeed be used to discover novel genetic associations via GWAS, which was not previously possible with traditional handcrafted IDPs such as volume, mass and function indices.

A schematic overview of the proposed UPE method is presented in Fig. \ref{fig:flowchart}. The details of each step are outlined in the Methods section. First, we extracted a surface mesh representation of the anatomical structures. In particular, we studied 3D meshes representing the LV at end-diastole from CMR images of the UK Biobank database using an automatic deep-learning-based segmentation method \citep{ref_xia2022segm}. We then learn a low-dimensional representation of the 3D meshes which captures anatomical variations using an encoder-decoder model. All meshes were projected onto this latent space to derive a few shape descriptors (or latent variables) for each of them. These features were finally used in GWAS for discovering genetic variants associated with shape patterns. Furthermore, to enhance discoverability, we adopt an ensemble-based approach: a set of phenotypes obtained through different models trained and configured with varying network metaparameters and weight initialisations (which induce diversity in the learned representations) are pooled together in one ensemble, yielding a redundant yet more expressive representation than the individual latent vectors. GWAS is performed against each phenotype of the ensemble, one at a time. A corrected Bonferroni threshold is then computed to declare the associations significant. We demonstrate that this approach is indeed effective in discovering additional biologically relevant genetic associations.

\section*{Results}


Convolutional mesh autoencoders were trained on the LV meshes at end-diastole, using a range of network metaparameters. For comparison, we also fit a shape PCA model (see Methods section).
The reconstruction performance for these models is displayed in Supplementary Figure \ref{fig:reconstruction_performance}.

GWAS was performed across all latent variables, for all training runs achieving a good reconstruction performance (see Methods section).
The number of such runs was $R=36$. Results obtained with $n_z=8$ and $n_z=16$ (eight and sixteen latent variables respectively) are reported, with a total number of 384 latent variables in the pooled representation. We also note that, despite the good reconstruction performance, shape PCA was not able to find genome-wide significant association when tested in GWAS (see Supplementary Figure \ref{fig:pca_gwas}).

Firstly, we examined the prevalence of the significant GWAS loci found, across runs. To count the loci, we split the genome into approximately LD-independent genomic regions \cite{ref_berisa2016approximately} and computed the number of loci below the usual genome-wide significance threshold of $5\times10^{-8}$ (see details in the Methods section); Table 1 
shows the results. Loci were annotated with gene names using the web tool FUMA (Functional Mapping and Annotation) \citep{ref_watanabe2017functional}. Among the candidate genes provided by this tool, a literature review was conducted in search for evidence of an association with cardiovascular phenotypes.

\subsection*{Genetic findings}
\label{subsec_GWAS}

\paragraph{Loci with study-wide significance.} We observe that the five most prevalent loci have prior evidence of association to cardiac phenotypes. PLN plays a crucial role in cardiomyocyte calcium handling by acting as a primary regulator of the SERCA protein (sarco/endoplasmic reticulum Ca$^{2+}$-ATPase), which transports calcium from the cytosol into the SR1 \citep{maclennan_2003}.Mutations in PLN have a well-established relationship with dilated cardiomyopathy (DCM) \citep{ref_Eijgenraam}.
In \citep{ref_pirruccello}, PLN was found to be associated with LVEDV and LVESV. However, \citep{ref_nayaung} does not report this locus for the same phenotypes.

The locus at chromosome 2 has been reported in  \citep{ref_nayaung, ref_pirruccello, ref_fractal_dim} and mapped to gene TTN. This gene encodes for protein titin, which is responsible for the sarcomere  assembly of the myocytes, and determines stretching, contraction and passive stiffness of the myocardium \citep{granzier_giant_2004}. 

The association with SNP rs4767239 is likely linked to gene TBX5 (T-box transcription factor 5), which has a known role in the development of the heart and the limbs \cite{ref_steimle2017tbx5}. Mutations in this gene have been associated, through familial studies, with Holt-Oram syndrome, a developmental disorder affecting the heart and upper limbs. Notably, it has not been reported on recent GWAS on LV phenotypes.

SNP rs35564079 is likely associated to gene NKX2.5. As it happens with TBX5, it plays a crucial role in heart development; in particular, in the formation of the heart tube, which is a structure that will eventually give rise to the heart and great vessels. NKX2.5 helps to determine the position of the heart in the chest and also plays a role in the development of the heart valves and septa. Mutations in the NKX2.5 gene have been linked to several types of congenital heart defects, including atrial septal defects and atrioventricular block \cite{ref_xu2017nkx25}. It has not been reported in \cite{ref_nayaung} or \cite{ref_pirruccello} but shows borderline significance with trabecular fractal dimension \cite{ref_fractal_dim}.

Interestingly, \cite{ref_fractal_dim} reports the GOSR2 locus as significantly associated to trabecular fractal dimension at slices 3 and 4, however previous GWAS on global LV indices have not reported this locus. 
More broadly in the literature on genetics of cardiovascular phenotypes, it has been reported as associated to ascending aorta distensibility \citep{ref_pirruccello2022deep_aorta}, mitral valve geometry \cite{ref_yu2022mitralvalve} and congenital heart disease \cite{lahm2021congenital}.

The association at SNP rs2245109 is likely linked to gene STRN, in chromosome 2. This gene codes for protein striatin, which is expressed in cardiomyocytes and it has been shown to interact with other proteins implicated in the mechanism of myocardial function \cite{ref_nader2017strn}. Moreover, mutations in this gene have been shown to lead to DCM in dogs. \cite{meurs2013association}. %

The locus near gene ATXN2 has been previously reported for LVEDV and LVSV. \cite{ref_pirruccello}

In addition to the loci with prior evidence discussed above, we report a number of novel genetic loci with $p < \PSW$, which have not been previously reported in connection with cardiac phenotypes. We note that all of these novel associations were detected in over 25\% of the runs (see Table 1). The loci were annotated based on the closest gene: CCDC91, LSP1, LGALS8 and EN1. We note that locus EN1 is also significant for LVEDSph (see Supplementary Figure \ref{fig:LVEDSph}). 

The GWAS summary statistics for the best latent variable of each of these 11 loci is displayed in Table 2.

\paragraph{Loci with suggestive significance.} In addition to genetic loci with $p<\PSW$, there is a number of SNPs with $\PSW<p<\PGW$, which we mention here by virtue of them showing up in more than 5 independent runs of the autoencoder. 

We first examine the loci close to genes with some known roles in cardiac physiology: genes BAG3
, WAC
, LMO7
, RBM20
and CNOT7. 
BAG3 is a gene coding for a cellular protein that is expressed predominantly in skeletal and cardiac muscle, which has a role in homeostasis of myocytes and in the development of heart failure \cite{ref_knezevic2015bag3}; also, it shows a strong association with LVESV, as found in previous studies \cite{ref_nayaung, ref_pirruccello} and in our own GWAS for this phenotype. 
The association near WAC is likely causally linked to the WAC gene; indeed, pathogenic deletions in this gene have been shown to lead to rare genetic disorders that produce, among other phenotypes, cardiac defects \cite{ref_quental2022wac}. However, further investigation would be needed to assess this hypothesis. 
Mutations in the LMO7 gene have been associated to Emery–Dreifuss muscular dystrophy, a disease with cardiac manifestations. Variants in the RBM20 are associated to DCM \cite{ref_koelemen2021rbm20}. 
Finally, CNOT7 codes for a protein, CCR4-NOT, which is a subunit of a protein complex called CCR4-NOT deadenylase whose activity was found to be required to prevent cardiomyocyte death in mice \cite{ref_yamaguchi2018ccr4}.

Three additional stable loci were found near genes CENPW, FILIP1L, and OR9Q1.

\subsection*{Effect on LV morphology.}
\label{morphology}
The effect of these loci on LV morphology was assessed by selecting the single phenotype with the strongest $p$-value for the associated locus. To help characterise these latent variables, Spearman correlation coefficient between the latter and LV handcrafted indices were computed and shown in Supplementary Table S3. 
These indices were LVEDV, LV sphericity index at end-diastole (LVEDSph), LV myocardial mass (LVM), and LV mass-to-volume ratio (LVMVR=LVM/LVEDV).

\begin{figure*}[ht!]
\centering
\includegraphics[width=\textwidth]{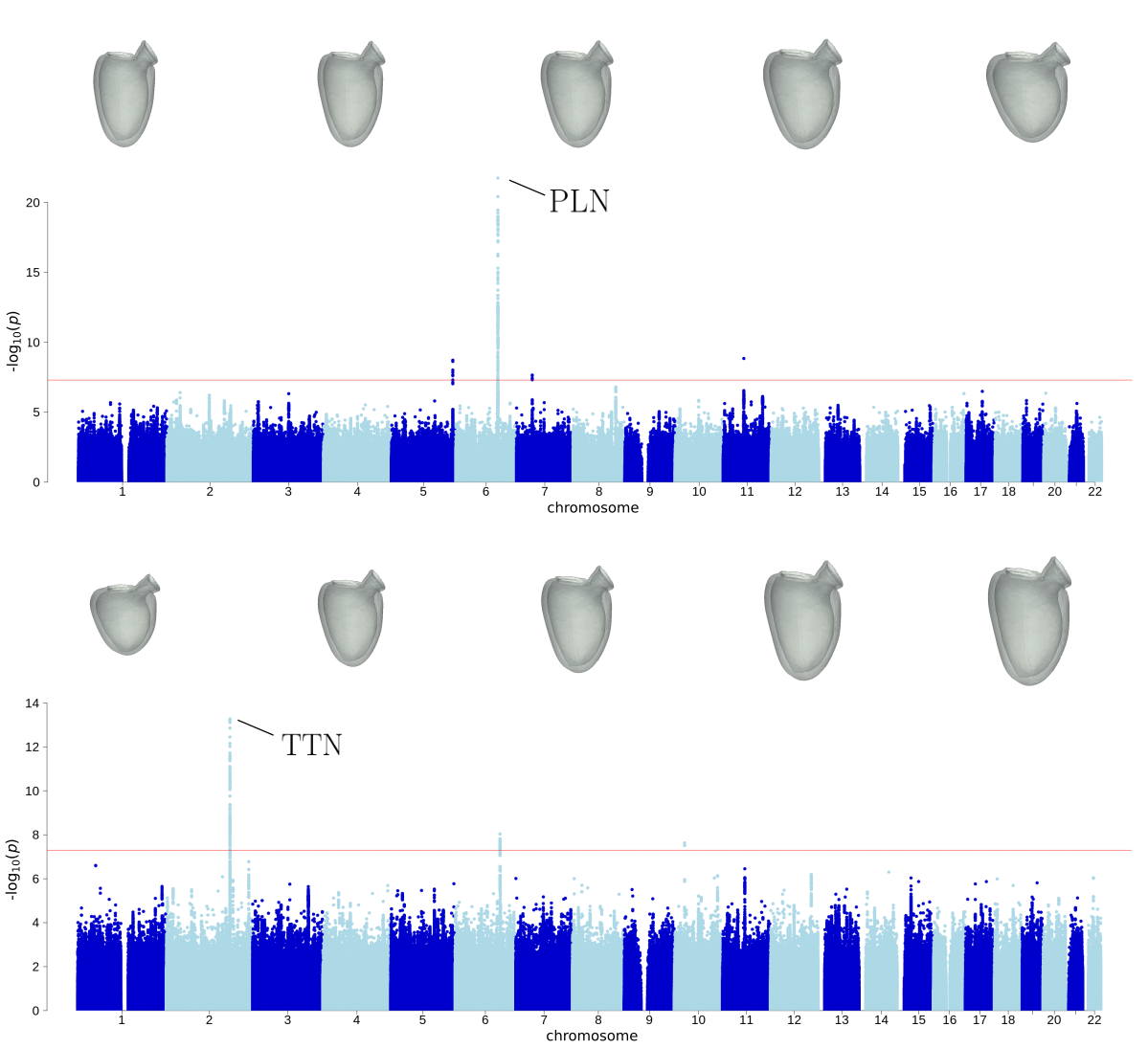}
\caption{Manhattan plots for LV latent variables with best association with a) PLN and b) TTN loci. On top are shown the average meshes corresponding to the following range of quantiles, for each latent variable: [0, 0.01], [0.095, 0.105], [0.495, 0.505], [0.895, 0.905] and [0.99, 1].
}
\label{fig:gwas}
\end{figure*}

We also examine the shapes of the average mesh for different ranges of quantiles for this latent variable, from 0 through 1. This is shown in Fig. \ref{fig:gwas}, along with the associated Manhattan plots, for loci PLN and TTN. For each of the remaining loci, the morphological effect for the latent variable with the strongest association can be found in the Supplementary Material, along with Manhattan plots and LocusZoom plots in a 1 Mb region centered at the locus.

Interestingly, we observe a very distinct effect on morphology for each of these loci. Whereas PLN variants influence a latent variable which is mainly linked to the sphericity (Spearman $r=0.625$) with a relatively small effect on LVEDV ($r=0.434$), gene TTN shows a greater correlation with the latter ($r=0.889$). Consistent with this, GWAS on LVEDSph shows no signal for TTN, but a strong one for PLN ($p=10^{-15}$, see Supplementary Figure \ref{fig:LVEDSph}), which is also in line with a previous finding of ours \cite{ref_bonazzola2021miccai}. 

On the other hand, gene TBX5 is linked to a latent variable which, as for TTN, is mainly correlated to LVEDV with no correlation to LVEDSph. For developmental gene NKX2.5 (see Supplementary Figure \ref{fig:chr5_103}), we note that the associated latent variable has an effect both in LVEDV ($r=0.865$) as in LVEDSph ($r=0.372$).
STRN locus is linked to a subtle phenotype controlling mitral orientation without a concomitant change in LV size (see Supplementary Figure \ref{fig:chr2_23}).

\begin{table*}[ht!]
\begin{center}
\begin{tabular}{ccrl}
\toprule
  region (hg37) &    locus name &  count &            minimum $p$-value \\
\midrule
 chr6:117672972-118963115 &\textbf{PLN} &     35 & $1.8 \times 10^{-22}$ \\
chr2:178553183-181312739 &\textbf{TTN} &     34 & $5.3 \times 10^{-14}$ \\
chr12:113986709-115036602 &     \textbf{TBX5} &     31 & $5.5 \times 10^{-12}$ \\
chr17:43056905-45876022 &\textbf{GOSR2} &     30 & $8.0 \times 10^{-15}$ \\
chr12:110336719-113263518 &  \textbf{ATXN2*} &     26 & $1.1 \times 10^{-11}$ \\
chr5:171074292-172678327 &  \textbf{NKX2.5} &     26 & $1.8 \times 10^{-11}$ \\
chr13:75670143-77410555 &  LMO7 &     23 & $6.6 \times 10^{-09}$ \\
chr6:125424383-127540461 &  CENPW* &     23 & $2.3 \times 10^{-10}$ \\
chr10:110317705-112561493 &    RBM20 &     21 & $4.3 \times 10^{-09}$ \\
chr11:1213590-3665481 &\textbf{LSP1*} &     20 & $1.6 \times 10^{-12}$ \\
chr8:15991660-17387876 &  CNOT7 &     17 & $7.4 \times 10^{-10}$ \\
chr10:26888684-29323236 &  WAC &     14 & $3.1 \times 10^{-09}$ \\ 
chr1:235819436-237555628 &\textbf{LGALS8*} &     14 & $1.2 \times 10^{-10}$ \\
chr12:27799773-29651255 & \textbf{CCDC91*} &     13 & $7.6 \times 10^{-13}$ \\
chr3:99373762-100592217 &  FILIP1L* &     12 & $1.0 \times 10^{-09}$ \\
chr2:118367466-121303783 & \textbf{EN1*} &     11 & $1.3 \times 10^{-10}$ \\
chr10:120591353-122407323 &  BAG3 &      8 & $8.5 \times 10^{-09}$ \\
chr11:55082657-58457495 &  OR9Q1 &      7 & $1.4 \times 10^{-09}$ \\
chr2:36122006-38132712 &  \textbf{STRN} &      6 & $7.0 \times 10^{-11}$ \\
\bottomrule
\end{tabular}
\caption{Counts of GWAS hits across runs, $\mathcal{C}_{\ell}$ for each locus $\ell$, which represents the number of runs for which the corresponding locus shows at least one association with $p< \PGW = \PGWVAL$ (see details in the Methods section). Note that the total number of runs was 36. Only regions with more than 5 counts are shown here. Genes with an asterisk were annotated based purely on proximity to the lead variant in that region. Gene names with no asterisk have additional prior evidence of a link to cardiac physiology. The last column represents the minimum $p$-value across the phenotype ensemble. Loci that surpass the study-wide threshold $\PSW (=\PSWVAL)$ are shown in bold letters.}
\end{center}
\label{table:gwas_counts}

\end{table*}

\begin{table*}[ht!]
\begin{center}
\begin{tabular}{llrrccr}
\toprule
locus name &   chromosome &           position (hg37) &                lead variant & $|\hat{\beta}|\pm\text{sd}(\hat{\beta})(\times 10^{-2})$ & $p$-value \\
\midrule
PLN & 6 &  118667522 &         rs11153730 &  $7.7 \pm 0.8$ &  $1.8\times 10^{-22}$ \\
GOSR2 &  17 &   45013271 &         rs17608766 & $9.2 \pm 1.1$ &  $3.9\times 10^{-16}$ \\
TTN &   2 &  179558366 &          rs2042995 & $7.2 \pm 1.0$ &  $5.3\times 10^{-14}$ \\
CCDC91* &  12 &   28544464 &          rs3741760 &  $6.6 \pm 0.9$ &  $7.6\times 10^{-13}$ \\
LSP1* & 11 &    1887068 &           rs569550 & $5.8 \pm  0.8$ &  $1.6\times 10^{-12}$ \\
TBX5 & 12 &  114816548 &          rs4767239 &  $7.3 \pm 1.1$ &  $5.6\times 10^{-12}$ \\
ATXN2* &  12 &  111907431 &         rs35350651 &  $5.4 \pm 0.8$ &  $1.1\times 10^{-11}$ \\
NKX2.5 &   5 &  172670611 &         rs35564079 & $6.0 \pm 0.9$ &  $1.8\times 10^{-11}$ \\
STRN &   2 &   37086197 &          rs2245109 & $5.2 \pm 0.8$ &  $7.0\times 10^{-11}$ \\
LGALS8* &   1 &  236691532 &          rs2853621 &  $5.5 \pm 0.9$ &  $1.2\times 10^{-10}$ \\
EN1* &   2 &  119479427 &           rs162748 & $5.5 \pm 0.9$ &  $1.3\times 10^{-10}$ \\
\bottomrule
\end{tabular}
\caption{11 genetic loci that surpass the study-wide significance threshold of $\PSW = \PSWVAL$, along with the summary statistics for the best phenotype for each locus. $\hat{\beta}$ and $\text{sd}(\hat{\beta})$ are the estimated effect sizes and their standard errors, respectively; these correspond to the inverse-rank-normalised phenotypes.}
\end{center}
\label{table:gwas_hits}
\end{table*}

\subsection*{Comparison with GWAS on traditional LV indices.}
\label{previous_gwas}
For comparison, we collected the GWAS summary statistics from previous studies on LV phenotypes, derived also from UKB CMR images, namely: \cite{ref_nayaung}, \cite{ref_pirruccello} and \cite{ref_fractal_dim}. We also include LVESV, LVSV and LVEF.
Note, however, that the unsupervised features studied in this work are static and were extracted using only the end-diastolic phase.

We also performed our own GWAS on traditional cardiac indices obtained using our segmentation approach (i.e. the indices were derived using the same meshes as the unsupervised phenotypes). Note that LVEDSph has not been investigated in previous GWAS, and is reported for the first time here, to the best of our knowledge (although a related phenotype, named "LV internal dimensions" was studied in an early GWAS of echocardiography-derived LV traits, \cite{ref_vasan2009genetic}). Details on how to compute this phenotype can be found in the Supplementary Material. 

\begin{figure*}[ht!]
\centering
\includegraphics[width=0.8\textwidth]{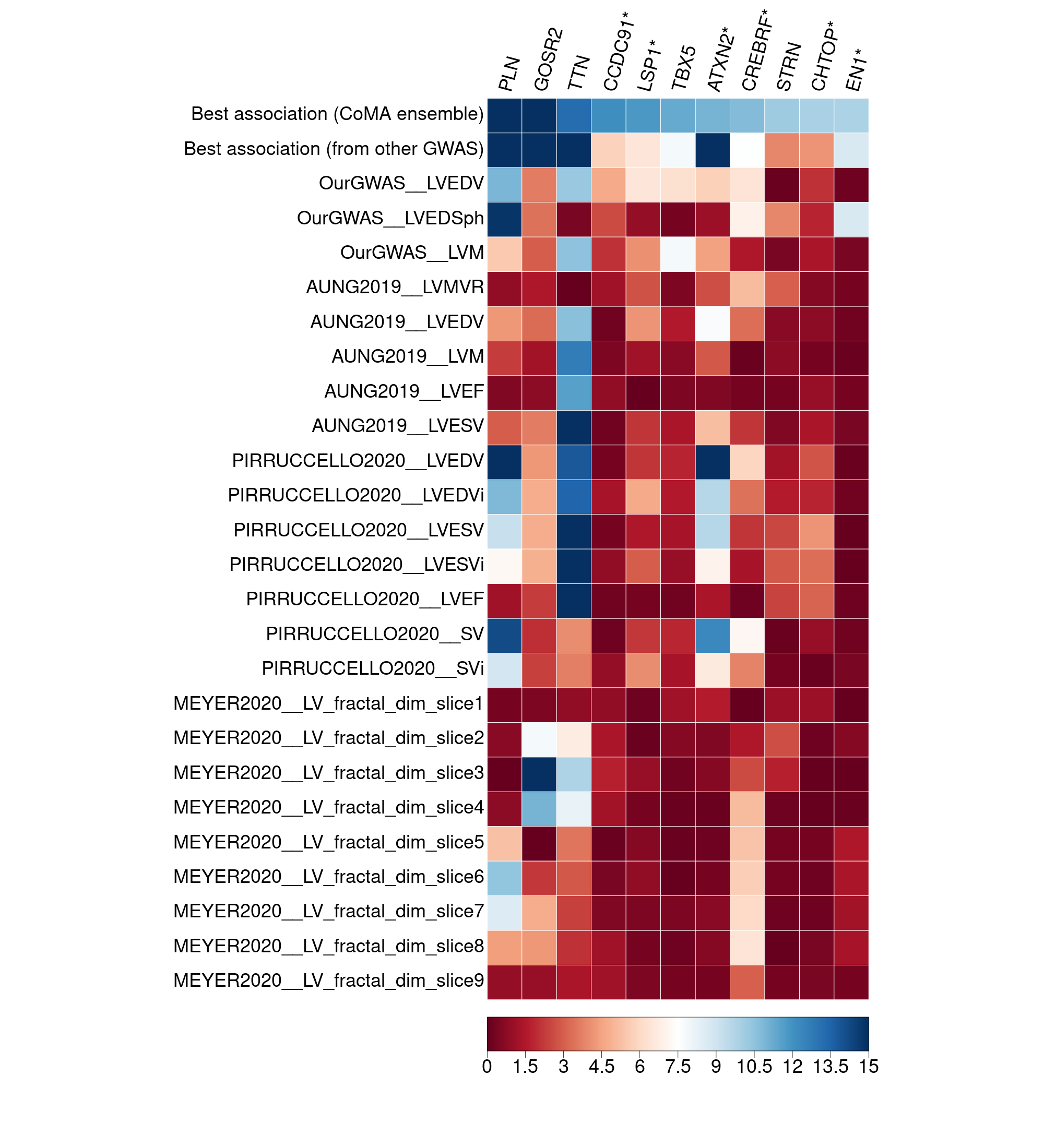}
\caption{Comparison of the $-\log_{10}(p)$ values for the 11 study-wide significant genetic loci found in this work, with GWAS on handcrafted cardiac indices. The top row corrsponds to the best association found for that locus across the ensemble of phenotypes, whereas the second row corresponds to the best $p$-value for that locus across the previous GWAS. White color corresponds to the genome-wide significance threshold of $5\times 10^{-8}$, whereas shades of red and blue correspond weaker and stronger associations, respectively. SV denotes stroke volume. LVEDVi, LVESVi and SVi denote the indexed versions of the phenotypes, i.e. the phenotype divided by the subject's body surface area.}
\label{fig:previous_gwas}
\end{figure*}

The comparison can be seen in figure \ref{fig:previous_gwas}. For each locus in Table 2 
(which all passed the Bonferroni threshold), this figure displays the association $p$-value found in previous GWAS. Shades of red represent non-genome-wide significant associations, whereas shades of blue represent genome-wide significant ones, and white corresponds to the $\PGW$ threshold. The second row represents the best $p$-value across all the traditional phenotypes for the loci given in the columns. 

\section*{Discussion}

As exposed in the Results section, we were able to retrieve study-wide significant loci that had been found in previous GWAS on handcrafted phenotypes (PLN, TTN, ATXN2 and GOSR2). In addition, genes with a known role in cardiac physiology (TBX5, NKX2.5, STRN) but no prior GWAS association, were identified with the same level of significance. Four additional loci constitute potential avenues for future research.
Finally, eight additional loci were found with suggestive significance ($\PSW < p <\PGW$ and $\mathcal{C}_{\ell}>5$). 5 of these have a prior evidence of a link to cardiac pathways. 

It is interesting to note that, for some loci, a relatively small number of runs produced a latent variable with a genome-wide significant association to the locus: the UPE approach seems to be crucial in order to retrieve this association, as it is likely to be missed in one individual autoencoder run.

Importantly, our approach allows to detect the milder effect on morphology of common variants near genes whose mutations are known to have highly deleterious effects (either by study of Mendelian diseases in humans, or by studies on model organisms). It is likely that these variants and the associated unsupervised LV features hold prognostic value, however this is uncertain at this point, and it should be possible to assess it once UKB releases more longitudinal data on the same subjects studied here.

As an interesting observation, we note that most of the phenotypes extracted here show a remarkable oligogenicity, i.e. they are controlled by few genes (see Figures \ref{fig:chr6_78} through \ref{fig:chr2_23}). This is in contrast to what happens with traditional indices where there are often many associations (see Supplementary Figures \ref{fig:LVEDV} through \ref{fig:LVMVR} or \cite{ref_nayaung, ref_pirruccello, ref_fractal_dim}). This suggests that the UPE approach allows to identify more optimal phenotypes for each locus, as compared to traditional handcrafted phenotyping.

In terms of gene discovery, the advantages of the ensemble-based phenotyping approach proposed here are best conveyed by examining the association $p$-values of the loci found in GWAS performed against traditional handcrafted phenoytpes, displayed in Figure \ref{fig:previous_gwas}. For example, when examining the GOSR2 locus, we found no genome-wide significant association when performing GWAS on traditional LV indices derived from the same meshes; neither have previous studies, with the exception of \cite{ref_fractal_dim} which investigated LV trabecular fractal dimension. Likewise, other genes, like STRN, which have prior knowledge of being implicated in cardiac pathways, have not been reported to date in mostly healthy cohorts such as UKB. Other examples are the loci near genes CCDC91 and LSP1, which achieve study-wide significance in our approach, however they have been reported in no previous GWAS on LV phenotypes (i.e. all squares are coloured with red shades). This highlights the shortcomings of traditional image-derived-phenotyping techniques when it comes to discoverability of relevant genes

In addition to improved discoverability, the UPE framework enables a better understanding of the genetic architecture of cardiac phenotypes, even for genetic loci that were known from previous studies. Most notably, TTN was shown to be linked to LV size, whereas PLN (which has been previously found in GWAS of LVEDV) controls a feature that jointly models changes in LV size and sphericity.

Based on our findings we argue that, in large-scale imaging studies it is crucial, along with increasing sample size, to count with good techniques to perform deep phenotyping that allows to boost gene discoverability in GWAS.

\section*{Conclusions}
In this work, we proposed a framework for left-ventricular (LV) phenotyping based on unsupervised geometric deep learning techniques on image-derived 3D meshes to discover genetic variations that affect LV shape through GWAS. The proposed methodology is based on finding a latent low-dimensional representation of the CMR-derived LV 3D meshes using convolutional mesh-autoencoders and then performing GWAS on the learned latent features. As hypothesised, this dimensionality reduction method, using Kullback-Leibler regularisation, yielded phenotypes with statistically significant genetic associations.  

The methodology of ensembling SNP associations across representations obtained through different network metaparameters, followed by the correction in the Bonferroni threshold necessary to control for false discover rate, has proven effective in identifying novel associations of mesh-derived phenotypes with genetic loci.
In addition to previously identified loci, namely TTN, PLN, TBX5, GOSR2 and ATXN2, we report 6 additional genetic loci that have not been discovered in recent GWAS of LV phenotypes. Moreover, we report a number of loci which do not surpass our corrected Bonferroni threshold, however their association remains suggestive by virtue of surpassing the usual genome-wide significance threshold of $p_{\text{GW}} =5\times10^{-8}$ in a number of phenotypes, obtained from independently trained autoencoder networks. Some of the latter, like BAG3, RBM20 and STRN, have been previously linked to cardiac pathways as have not been previously linked to cardiac physiology.

We argue that the proposed ensembling approach is not only useful for discovering novel associations, but also enables a deeper understanding of the effect of previously known genes: indeed, the effect of the latent variables with the strongest association $p$-values for each locus can be used as suggestive evidence of the role of that locus in LV shape. For example, we found that TTN and PLN, which had been found in previous GWAS to correlate with LV volume, have actually a distinct effect on LV shape. Whereas TTN shows indeed a clear effect on LV size, PLN is linked to a more complex phenotype which mainly involves changes in LV sphericity.

More generally, these results validate our methodology to extract knowledge about the genetics driving the morphology of organs, leveraging databases that provide linked genetic and imaging data, such as the UKB. This methodology can be used seamlessly to study surface meshes of other organs, like the brain or the skull \citep{ref_roosenboom2018skull, ref_fan2022brain}. Additionally, the algorithm proposed here can be extended to process 3D cardiac meshes throughout the cardiac cycle to capture anatomy as well as quantitative features related to patterns of contraction and relaxation. Future studies will explore these directions.

\section*{Methods}
The proposed method is outlined in Figure \ref{fig:flowchart}. It starts with the extraction of 3D meshes representing the LV from CMR images using an automatic segmentation method \citep{ref_xia2022segm}.  
 We then train several models with different metaparameters (network architecture, random seeds controlling weight initialization and dataset partitioning, and relative weight of the variational loss) to learn low-dimensional representations of the 3D meshes which capture anatomical variations using an encoder-decoder model. All meshes are then projected to this latent space to derive a few shape descriptors (or latent variables) for each mesh. In order to take advantage of the variability induced in the representation obtained by the metaparameters, we pooled the different latent vectors together to obtain a richer representation. The features that constitute this pooled representation are finally used in GWAS to discover genetic variants associated with shape patterns. 

\subsection*{Description of the data}
The proposed framework can discover novel associations between genetic variations and morphological changes in anatomical structures. We showcase its potential in the context of cardiac images acquired within the UK Biobank (UKB) project (data accession number \ACCESSIONNUMBER). The UKB is a prospective cohort study that between 2006 and 2010 recruited around half a million volunteers across the United Kingdom, aged 40-69 years old at the time of recruitment \citep{ref_ukbb}. The project collected a vast amount of phenotypic information about its participants and linked them to their electronic health records (EHR). The collected data includes, among others, genetic data from single-nucleotide polymorphism (SNP) microarrays for all the individuals, and also CMR data for a subset of them (which comprises over 50,000 individuals at the moment of this writing, but is planned to reach 100,000). These datasets are described in \citep{ref_ukbb_genetics} and \citep{ref_ukbb_cmr}, respectively.

\subsubsection*{CMR data}
The CMR imaging protocol used to obtain the raw imaging data is described in \citep{ref_ukbb_cmr}. 
We employed an automatic segmentation method \citep{ref_rahman} to segment the LV in the CMR images. This method generates a set of registered 3D meshes, i.e. meshes with the same number of vertices with consistent identical connectivity between them. There is one mesh per subject and per time point. In this work, we only use the LV mesh at end-diastole. The LV mesh for subject $i$, $i=1,...,N$, can then be represented as pairs $(\textbf{S}_i, A)$, where $\textbf{S}_i=\big[\,x_{i1}\,y_{i1}\,z_{i1}\,|\,...\,|\,x_{iM}\,y_{iM}\,z_{iM}\,\big]\in \mathbb{R}^{M\times 3}$ is the shape and $A$ is the adjacency matrix of the mesh. We also define, for convenience, the vectorised form of the shapes $\textbf{s}_i=\big(x_{i1},y_{i1},z_{i1},...,x_{iM},y_{iM},z_{iM}\big)\in \mathbb{R}^{3M}$. The adjacency matrix is such that $A_{jk}=1$ if and only if there is an edge between vertices $j$ and $k$ and $A_{jk}=0$ otherwise. The cardiac meshes also have the property of being triangular and closed, so $A_{jk}=A_{kl}=1\implies A_{jl}=1$ for all vertices $i$, $j$ and $k$.

\subsubsection*{Genotype data}
SNP microarray data is available for all the individuals in the UKB cohort. This microarray covers \NCALLS\, genetic variants including SNPs and short insertions and deletions. 
The SNP microarrays used in UKB have been described in \citep{ref_ukbb_genetics}. From these genotyped markers, an augmented set of over 90 million variants was imputed. The GWAS was performed across the latter dataset, particularly on the autosomes (chromosomes 1 through 22).

The usual quality control steps on the genetic data were performed. This included filtering out rare variants using a threshold for MAF of 1\% (within the subcohort of \NCMRGBR subjects), Hardy-Weinberg equilibrium $p$-value less than $10^{-5}$ and low imputation information score (less than 0.3). This results in a set of \NIMP genetic variants. 

\subsection*{Unsupervised representation learning for genetic discovery}
\label{results:dimensionality_reduction}
Given the set of meshes representing the anatomical structure of interest (LV meshes), the pose-sensitive parameters (translation and rotation) were removed using generalised Procrustes analysis.
\begin{equation}
\bar{\textbf{S}}=\frac{1}{N}\sum_{i=1}^{N}{\textbf{S}},
\end{equation}

\begin{equation}
\bar{\textbf{s}}=\frac{1}{N}\sum_{i=1}^{N}{\textbf{s}},
\end{equation}

\begin{equation}
\textbf{C}=\frac{1}{N-1}\sum_{i=1}^{N}({\textbf{s}}_i-\bar{\textbf{s}})({\textbf{s}}_i-\bar{\textbf{s}})^t.
\end{equation}

Here we propose to learn a reduced set of features that best describe cardiac shape using convolutional mesh-autoencoders (CoMA). We will compare the proposed approach with the well-known method of principal component analysis (PCA). While in PCA only vectorised 3D point clouds $\textbf{s}_i$ will be provided as input (therefore ignoring the data structure and topology), convolutional mesh-autoencoders leverage topological information about the connectivity between the vertices for learning more powerful non-linear representations. However, both approaches can be thought of as particular cases of the encoder-decoder paradigm.

In such a paradigm, there is a pair of encoding and decoding functions, $E_{\theta}:\mathbb{R}^{3M}\rightarrow\mathbb{R}^{n_z}$ and $D_{\phi}:\mathbb{R}^{n_z}\rightarrow\mathbb{R}^{3M}$ that are parameterised by a set of learnable coefficients $\theta$ and $\phi$, respectively. $n_z\in\mathbb{N}$ is the size of the latent space, and it is usually chosen so that $n_z\ll M$ (hence the dimensionality reduction). 

Optimal parameters $\theta^*$ and $\phi^*$ for reconstruction can be estimated by making the composite function $D_{\phi} \circ E_{\theta}$ as close to the identity function $I$ as possible over the training set $\mathbb{S}_\text{train}\subset\mathbb{S}$, using some reasonable measure of reconstruction error $L_{\text{rec}}$ (examples of which are the $L_1$ norm, the $L_2$ norm or the mean squared error MSE) along with a regularisation term $\Omega$, which will account for additional constraints we want to impose on the model. We want to minimise the following function with respect to $\phi$ and $\theta$: 

\begin{equation}
L(\mathbb{S}_\text{train}|\theta, \phi)=
L_{\text{rec}}(\mathbb{S}_\text{train}|\theta, \phi)+
\beta\Omega(\mathbb{S}_\text{train}|\theta, \phi).
\label{eq_loss_function}
\end{equation}

\noindent where $\beta\in\mathbb{R_{\geq 0}}$ is a weighting coefficient for the regularisation term. $\textbf{z}_i:= E_{\theta^*}  (\textbf{S}_i)\in\mathbb{R}^{n_z}$ would then be a low-dimensional representation of the shape $\textbf{S}_i$, whereas $\hat{\textbf{S}}_i:=\big(D_{\phi^*} \circ E_{\theta^*}\big)(\textbf{S}_i)$ is the associated reconstructed shape.

\subsubsection*{Principal component analysis.}
PCA is a standard linear technique for dimensionality reduction \citep{pearson_pca}. In terms of the encoder-decoder framework detailed above, it can be obtained by requiring $D$ and $E$ to be linear transformations and using the $L_2$ norm, besides imposing an orthogonality constraint on the latent vectors \citep{goodfellow-et-al-2016}.

The idea is to find a basis of vectors  $\mathcal{B}_{n_z}=\{\textbf{v}_i\}_{i=1}^{n_z}\subset\mathbb{R}^{3M}$
for a fixed $n_z < 3M$. The $n_z$-dimensional linear subspace generated by $\mathcal{B}_{n_z}$ captures as much of the data variability as possible. It can be shown that such a basis corresponds to the $n_z$ eigenvectors of  $\textbf{C}$ with the largest eigenvalues; i.e. if $\textbf{C}=U^{t}\Lambda U$ where $\Lambda_{ij}=\delta_{ij}\lambda_i$ and $\lambda_i \geq \lambda_j$ if $i\leq j$, then $\mathcal{B}_{n_z}=\{{U\textbf{e}_i}\}_{i=1}^{n_z}$.
$\delta_{ij}$ is the Kronecker delta, which equals 1 if $i=j$ and 0 otherwise.

\subsubsection*{Convolutional mesh autoencoder}
In an autoencoder, both the encoding and decoding functions are feed-forward neural networks.
Inspired by recent works on unsupervised geometric deep learning \citep{ref_coma} for facial meshes, we propose constructing a convolutional mesh-autoencoder which uses spectral convolutions \citep{ref_spectral_graph_conv} to learn non-linear and low-dimensional representations of cardiac mesh structures. Here each layer of the encoder and decoder implements convolution operations parameterised by the graph Laplacian, to leverage information about the local context of each vertex. In order to learn global features, a hierarchical approach is used where each layer of the encoder and decoder implements downsampling and upsampling operations, respectively. 
Since the vertices are not in a rectangular grid, the usual convolution, pooling and unpooling operations defined for such topology (usually employed in image analysis) are not adequate for this task and need to be suitably adapted. Several methods have been proposed to do this \citep{ref_bronstein_geom_DL} which can be mainly classified in two broad groups: spatial or spectral. The approach proposed in this work belongs to the latter category, which relies on expressing the features in the Fourier basis of the graph, as explained below.

\paragraph{Spectral convolutions.} The Laplace-Beltrami operator $\mathcal{L}$ (or, more simply, the Laplacian) of a graph with adjacency matrix $A$ is defined as $\mathcal{L}:=D-A$, where $D$ is the degree matrix, i.e. a diagonal matrix with $D_{ii}:=\sum_{j}A_{ij}$ being the number of edges that connect to vertex $i$. The Fourier basis of the graph can be obtained by diagonalising the Laplace operator, $\mathcal{L}=U^t\Lambda U$. The columns of $U$ constitute the Fourier basis, and the operation of convolution $\star$ for a graph can be defined in the following manner:

\begin{equation}
x\star y :=U(U^tx\odot U^ty),
\end{equation}{}

\noindent where $\odot$ is the element-wise product (also known as Hadamard product), and $x$ and $y$ are arbitrary functions defined over the vertices of the graph. Spectral methods rely on this definition of convolution and differ from one another in the specific filter used. In this work, a parameterisation proposed in \citep{ref_spectral_graph_conv} will be used. 
The said method is based on the Chebyshev family of polynomials $\{T_i\}$. The kernel $g_\xi$ is defined as:

\begin{equation}
g_{\xi}(\mathcal{L})=\sum_{i=1}^{K}\xi_i T_i(\mathcal{L}).
\label{eq_chebyshev_filter}
\end{equation}

\noindent $K$ is the highest degree of the polynomials considered (in this work $K=6$). Chebyshev polynomials have the advantage of being computable recursively through the relation $T_i(x)=xT_{i-1}(x)-T_{i-2}(x)$ and the base cases $T_1(x)=1$ and $T_2(x)=x$. It is also worth mentioning that the filter described by Equation \ref{eq_chebyshev_filter}, despite its spectral formulation, has the characteristic of being local.

\paragraph{Autoencoder.} The downsampling and upsampling operations used in this study were proposed in \citep{ref_coma} based on a surface simplification algorithm proposed in \citep{ref_quadric_error}. These operations are defined before training each layer, using a single template shape. Here we utilise the mean shape $\mathbf{\bar{S}}$ as a template.

In each layer of the encoder, the downsampling operation generates a new triangular mesh (with its corresponding new Laplacian), such that the quadric error is minimised. The upsampling operations are created while performing the downsampling: the coordinates of the decimated vertices with respect to the remaining vertices are stored for each layer. 

\paragraph{Variational Autoencoder.} A Kullback-Leibler (KL) divergence term was added to encourage statistical independence of the different components of the latent representation, which is expected to improve its interpretability \citep{ref_betavae}. We hypothesise that it will also contribute to producing features with higher heritability, i.e. suitable candidate phenotypes to perform GWAS on.

To train a model with such a loss function, the framework of variational autoencoder (VAE) is utilised. In this framework, during the training phase the encoder maps the input into a probability distribution instead of a fixed vector. To emphasise this, we will replace the notation $E_\theta(\textbf{S})$ for the encoder network with $q_{\theta}(\textbf{Z}|\textbf{S})$, where $\textbf{Z}$ is now a random variable. During training, for the $j$-th latent variable (with $1\leq z_j\leq n_z$) two quantities are learned, $\mu_j$ and $\sigma_j$, and a realisation $z_j$ of the random variable $Z_j\sim\mathcal{N}(\mu_j, \sigma^2_j)$ is produced and passed through the decoder to generate the output mesh. The aforementioned KL-divergence term is then used to encourage the variational approximate posterior to be a multivariate Gaussian with a diagonal covariance structure. The regularisation term is computed as:

\begin{align}
\begin{split}
\Omega(\mathbb{S}_\text{train}|\theta, \phi)&= \mathbb{E}_{\mathbf{s}\sim\hat{p}_{\text{train}}}\
D_{\text{KL}}\Big(q_{\theta}(\textbf{Z}|\textbf{S})||\mathcal{N}(\mathbf{Z};\mathbf{0}, \mathbb{1}_{n_z})\Big)\\
&=\mathbb{E}_{\mathbf{s}\sim\hat{p}_{\text{train}}}
\frac{-1}{2n_z}\sum_{j=1}^{n_z}\Big(\log\sigma^2_j-\sigma^2_j-\mu^2_j+1\Big)
\label{eq_regularisation}
\end{split}
\end{align}

\noindent where $\mathbb{1}_{n}$ is the $n\times n$ identity matrix, $D_{\text{KL}}(p||q)$ is the KL divergence between probability distributions $p$ and $q$, and $\hat{p}_{\text{train}}$ is the empirical probability distribution associated with $\mathbb{S}_\text{train}$.
$D_{\text{KL}}(p||q):=\int p(x)\ln{\frac{p(x)}{q(x)}}dp(x)$. The last equality in Equation \ref{eq_regularisation} arises
from the formula for the KL divergence between two normal distributions where the second one is also standardised. During testing, the mode of the latent distribution, $\pmb{\mu}(\textbf{S})$, is the latent representation of the shape $\textbf{s}$. In the following, we will rename the weighting coefficient $\beta$ from Equation \ref{eq_loss_function} as $w_{\text{KL}}$ to make it more memorable.

\subsection*{GWAS}
According to the traditional GWAS scheme, we tested each genetic variant, $X_i\in\{0,1,2\}$, for association with each latent variable $z_k$ through a univariate linear additive model of genetic effects:

\begin{equation}
z_k = \beta_{ik}X_i+\epsilon_{ik}
\label{eq_gwas}
\end{equation}
\noindent where $\epsilon_{ik}$ is the component not explained by the genotype, assumed to be normally distributed. The null hypothesis tested is that $\beta_{ik}=0$. 

Only unrelated individuals with self-reported British ancestry were included in the study, to avoid issues related to population stratification. This produced a sample size of \NCMRGBR individuals. Summary statistics of demographic data from these subsample can be found in Supplementary Table S1.
For the results presented in the main text, no individuals were filtered out based on previous diagnoses or image-derived cardiac function parameters (such as ejection fraction).

Before GWAS, the phenotypes (i.e. latent variables) were adjusted for a set of covariates: sex, age, height, weight, body-mass index, body surface area, systolic and diastolic blood pressure, alcohol consumption, smoking status and the 10 top genomic principal components (computed within the British population). Details on how to compute the genomic principal component loadings, as well as on the pre-processing of demographic data, are provided in the Supplementary Material. To make this adjustment, multivariate linear regression on these covariates was performed, and then the residues of this regression were rank-inverse-normalised. These inverse-normalised residues are the phenotypic scores to be tested in the GWAS.

It is worth mentioning that the GWAS is performed on all the individuals, including those on which the dimensionality reduction algorithm was trained. This is correct because the algorithm does not optimise for association with genetic variants, and therefore a uniform distribution of $p$-values under the null distribution can be safely assumed even when including these subjects in the sample.

\subsection*{Unsupervised phenotype ensemble (UPE)}
Given that the evaluation metric which guides the training, i.e. reconstruction error with a variational loss,  is not necessarily aligned with the final purpose of discovering genes that influence LV shape, there is no reason to adopt the single run with the best value for such metric. This is the approached followed in our previous work, \cite{ref_bonazzola2021miccai}. Indeed, the observation that a number of loci are detected in only a small subset of runs indicates that following such a procedure would lead to failure to discover a number of relevant genetic loci. For this reason, here we propose to adopt an ensemble-based approach, in which we pool the different phenotypes together in a redundant yet more expressive representation. Based on the observation that different network metaparameters, dataset partitioning and weight initialisations yielded latent representations with different genome-wide-significant loci, we proposed to build an ensemble of phenotypes by concatenating the latent vectors for each individual run. This composite representation provides a redundant yet more expressive representation of LV shape at end-diastole. These runs covered a wide range of $w_\text{KL}$, and variations in network architectures; most importantly, in the latent dimension, $n_z$. Also, for a given combination of metaparameters (including architecture), an optimal learning rate was found and then 5 different random seeds were utilised to initialize the network's weights and to partition the full dataset into train, validation and test sets (each seed constitutes a different run). Details on the architectural parameters are given in the Supplementary Table S2.

From the full set of runs, we selected \NRUNS training runs that achieved a good reconstruction performance: a root mean squared deviation (RMSD) of less than \RMSETHR (averaged over the subjects from the test set). The deviation is taken to be the vertex-wise Euclidean distance, and the mean is taken over the 5220 vertices of the LV mesh. In other words, the RMSD for subject $i$ in run $r$ is: 

\begin{equation}
\text{RMSD}_{i,r}=\sum_{j=1}^{5220}\sqrt{||\textbf{x}_{i,j} - \hat{\textbf{x}}_{i,j}^{(r)}||_2^2},
\end{equation}

\noindent where $\textbf{x}_{i,j}$ denotes the triad of spatial coordinates for vertex $j$ in the mesh of subject $i$, and $\hat{\textbf{x}}_{i,j}^{(r)}$ is the same for the reconstructed mesh in run $r$ of the autoencoder. $||\cdot||_2^2$ denotes the squared Euclidean norm. Importantly, the runs were selected based \emph{only} on mesh reconstruction error and not in the presence or absence of GWAS hits. This allows to assume a uniform distribution of $p$-values over the $[0,1]$ interval.

These \NRUNS runs produced a total of \NZVAR  phenotypes (where the latent dimension was 8 for some runs and 16 for others). 
In order to control for the false discovery rate, this procedure requires correcting the usual genome-wide Bonferroni $p$-value threshold, $\PGW = 5\times 10^{-8}$, since the number of statistical tests that are performed grows with the size of the (pooled) representation. In order not to overcorrect this threshold, whenever a pair of latent variables (within the same run or not) had a Spearman correlation coefficient greater than 0.95 in absolute value, one of them was dropped at random. This procedure resulted in \NZVARPRUNED phenotypes to be tested in GWAS. The new study-wide threshold utilised was, therefore,  $p_{\text{SW}}=\frac{p_{\text{GW}}}{324}=1.5\times 10^{-10}$.

Given that for each genomic locus, the lead variant might vary across different phenotypes by virtue of high linkage disequilibrium with close genetic variants, we adopt the following approach for locus counting: the genome is partitioned into 1703 approximately LD-independent regions, where each is region is nearly 2 megabases (Mb) in length. We compute the number of autoencoder runs in which each region $\ell$ was genome-wide significant, denoting this quantity $\mathcal{C}_{\ell}$: for each run $r$ and region $\ell$, we retrieve the minimum $p$-value,  across the different latent variables $z_k^{(r)}$ (remember that $1 \leq k \leq 8$ or $1 \leq k \leq 16$, depending on the run) which we call $p_{\ell,r}$. We then count the number of runs for which $p_{\ell,r}<\PGW$: $\mathcal{C}_{\ell}=\sum_{k=1}^{R}\mathbf{1}_{p_{\ell,r} <  p_{\text{GW}}}$, where $\mathbf{1}$ denotes the indicator function and $R=\NRUNS$. This $\mathcal{C}_{\ell}$ is the quantity labelled 'count' in Table 1.

{\small
\bibliography{gwas_cardiomorpho.bib}
}

\section*{Acknowledgements}

This project was funded by the following institutions: The Royal Academy of Engineering [grant number  CiET1819\textbackslash 19], EPSRC [TUSCA, grant number EP/V04799X/1] (R.B., N.R. and A.F.F.), The Royal Society, through the International Exchanges 2020 Round 2 scheme [grant number IES\textbackslash R2\textbackslash 202165] (R.B., E.F. and A.F.F).
E.F. was also funded by ANPCyT [grant number  PICT2018-3907] and UNL [grant numbers CAI+D 50220140100-084LI, 50620190100-145LI] (E.F).
We would like to thank Andres Diaz-Pinto, Peter Claes, Rahman Attar, Francisco Ibarrola and Sadaf Raza for useful discussions as well as editorial reviews on the manuscript.
This research has been conducted using data from UKB, a major biomedical database. We would like to thank the participants and members of the staff for enabling this research.
This work was undertaken on ARC4, part of the High Performance Computing facilities at the University of Leeds, UK.

\section*{Author contributions statement}
R.B. Design of the work, implementation, data analysis and interpretation, drafting the article. \newline
E.F. Design of the work, drafting the article, critical revision of the article, supervision of the project.\newline
N.R. Drafting the article, supervision of the project.\newline
Y.X. Provision of input data, critical revision of the article.\newline
B.K, Critical revision of the article, data interpretation. \newline
S.P, Critical revision of the article, data interpretation. \newline
T.S.M. Critical revision of the article, supervision of the project.\newline
A.F. Conception of the work, drafting the article, critical revision of the article, supervision of the project.\newline
All authors reviewed the manuscript.

\section*{Additional information}

\textbf{UK Biobank Accession code:} 11350. \newline
\textbf{Competing interests:} The authors declared no competing interests related to this work.

\pagebreak

\begin{center}
\textbf{\large Supplemental Materials: Unsupervised ensemble-based phenotyping helps enhance the discoverability of genes related to heart morphology}
\end{center}
\setcounter{equation}{0}
\setcounter{figure}{0}
\setcounter{table}{0}
\makeatletter
\renewcommand{\tablename}{Supplementary Table}
\renewcommand{\figurename}{Supplementary Figure}
\renewcommand{\theequation}{S\arabic{equation}}
\renewcommand{\thefigure}{S\arabic{figure}}
\renewcommand{\thetable}{S\arabic{table}}
\def\code#1{\texttt{#1}} 

\section*{Data and code availability}

\subsection*{Data}
The genetic and imaging data used for this work have been downloaded from the UK Biobank (UKB), through our application number 11350 and have not been included in this submission as they are protected data for which access must be granted by the UKB.
Cardiac magnetic resonance (CMR) images were processed using our own pipeline, whose output are 3D meshes for each subject and time point.
Summary statistics for the demographic variables of the cohort used in this study are shown in Table S1.

\subsection*{Code availability}
The codebase for this work is hosted at the GitHub repository \url{www.github.com/cistib/CardiacGWAS}. We will refer to this as the root repository. It contains two Git submodules: \url{www.github.com/cistib/CardiacCOMA} and \url{www.github.com/cistib/GWAS-pipeline}. Please note that the latter repository is public, however the rest of them are private at the moment. If you wish to access this code please contact the first author of this work at \code{scrb@leeds.ac.uk}.

The \code{CardiacCOMA} repository the code for an implementation of the convolutional mesh autoencoders (CoMA), in PyTorch and PyTorch Lightning. Hyperparameters and metrics are logged using the MLflow Python API. It also contains a Dockerfile to reproduce the software environment necessary to train the network, as a Docker container.

The second submodule contains the code to perform data pre-processing for GWAS, GWAS execution and results visualization. This repository is written in R and Python.

The root repository (\code{CardiacGWAS}) also contains code to produce the different figures of the paper, and to produce LaTex code for the different tables of results.
Finally, it contains a folder called \code{shiny} with source code of a R Shiny web application that allows to explore extensively the full set of results generated for this work (see following subsection).

\subsection*{Shiny app}
In order to confer more transparency to this work, a Shiny app is served which can be queried to examine the details of the full set of runs performed for this paper. It also allows to query additional data that do not contribute to the core message of this paper including, among other things, associations of the different latent variables to non-image data and to disease status, and detailed hyperparameter data on each autoencoder run. Details on how to connect to this app can be found at \url{www.github.com/rbonazzola/LV-GWAS-ShinyApp.git}.

\begin{table}[ht!]
\begin{center}
\begin{tabular}{|c|c|}
\hline
Male proportion & 48.29\% \\ \hline
Age (years) & $59.3\pm14.9$ \\ \hline
Height in males (cm) & $176.4\pm13.1$ \\ \hline
Height females (cm) & $163.1\pm12.2$ \\ \hline
BMI in males (kg/m$^2$)& $27.1\pm7.5$ \\ \hline
BMI in females (kg/m$^2$)& $26.1\pm9.0$ \\ \hline
SBP in males (mmHg) & $140.2\pm33.9$ \\ \hline
SBP in females (mmHg) & $133.5\pm30.0$\\ \hline
DBP in males (mmHg) & $82.4\pm17.1$\\ \hline
DBP in females (mmHg) & $78.4\pm17.6$ \\ \hline
\end{tabular}
\end{center}
\caption{Summary statistics of the demographic variables of the subsample of \NCMRGBR unrelated British individuals used in this work. Continuous variables are expressed as mean $\pm$ 2 s.d.}
\label{table:demographic_data}
\end{table}

\subsection*{Image and mesh processing}
Images were processed through a segmentation pipeline based on the deep learning approach in \cite{ref_xia2022segm}. This network was trained on a dataset of manually segmented images at end-diastole and end-systole (4700 subjects), for which the manual 2D contours were registered to a 3D cardiac atlas encompassing the 4 chambers \cite{ref_rodero2021atlas}, producing a 3D mesh for each subject and time point. This set of meshes was used to fit a point distribution model (PDM), consisting of 70 principal components. Briefly, the segmentation approach takes as input a stack of short axis (SAX) views, the different longitudinal axis (LAX) slices and participant metadata, and produces the 70 PCA loadings that allow to reconstruct the mesh for that subject.

It is worth mentioning that the cardiac atlas has a very high resolution, with 194541 nodes for whole heart and 52193 for the left ventricle. Since this makes them too large to be stored permanently for the whole population, they were decimated to 10\% and subsetted for the LV, producing the final size of $M=5220$ nodes. To ensure that all the decimated meshes had the same connectivity and their vertices were in correspondence (a requirement of the dimensionality reduction approach), a suitable decimation approach was used, namely the quadric error minimisation approach \cite{ref_quadric_error}, which is also used in the pooling layers of the mesh autoencoder. The code for computation of the downsampling matrices can be found at \code{CardiacCOMA/utils/CardioMesh}.

\subsection*{Genotype pre-processing}

The pre-processing of the imputed genotypes, consisting of filtering for individuals and genetic variants, was performed using \code{qctool} (v2.0.8). GWAS were performed using the BGENIE tool (version 1.4), and executions were conducted in the ARC4 HPC at University of Leeds, using batch mode. The runs were parallelized using the \code{Son of Grid Engine} (SGE) job scheduler available in the HPC, and Python scripts to create and submit bash jobs to the queue.
This filtering, as described in the main text, included excluding rare variants (MAF $<$ 1\% within the subcohort of \NCMRGBR British subjects), Hardy-Weinberg equilibrium $p$-value less than $10^{-5}$ and low imputation information score (less than 0.3). This resulted in a set of \NIMP genetic variants.

\paragraph{Genome partitioning.} In order to perform locus counting (as detailed in the Methods section), a set of nearly LD-independent regions of around 2Mb were utilised to partition the genome \cite{ref_berisa2016approximately}. The file for these regions is located at \code{GWAS\_pipeline/data/ld\_indep\_regions/fourier\_ls-all\_EUR\_hg19.bed}.

\subsection*{Covariates}
In this subsection, we describe the covariates that were used to adjust the phenotypes tested in GWAS. Code to perform the adjustment for covariates can be found at \code{GWAS\_pipeline/utils/preprocessing\_for\_gwas\_helpers.R}.

\paragraph{Demographic variables.} As described in the main text, the demographic variables used as covariates were: height, genetic sex, age, BMI, body surface area (BSA), systolic blood pressure (SBP), diastolic blood pressure (DBP). The respective field codes from the UKBB were 50 (``standing height``), 22001 (``genetic sex``), 21003 (``age when attended assessment centre``, instance 2 corresponding to the first imaging visit), 21001 (``body mass index``), 4080 (``systolic blood pressure``), 4079 (``diastolic blood pressure``). BSA was estimated as $0.20247\times (\text{weight}^{0.425}) \times (\text{height}^{0.725})$, where the weight is expressed in kg and height is expressed in meters. SBP and DBP were adjusted for those participants taking blood-pressure-lowering effect from the verbal interview (field 20003); SBP was adjusted by adding 15 mmHg, whereas DBP was adjusted by adding 10 mmHg (following the procedure in \cite{ref_nayaung}). Smoking status was categorised into ‘Current’, ‘Previous’ or ‘Never’, according to field 20116. Regular alcohol use was defined as a binary variable based on whether the participant reported consumption of alcohol at least three times per week (field 1558). Code for preprocessing the UKB files can be found at \code{GWAS\_pipeline/utils/demographics\_for\_gwas.R}.

\paragraph{Genetic principal components.} To compute the genomic PCA loadings, a similar approach as detailed in the UKBB genotyping QC report guide was used. It is reproduced here for convenience:
\begin{itemize}
 \item Minor allele frequency $\geq$ 2.5\% and missingness $\leq$ 1.5\%. (Checking that HWE holds in a subset of samples with European descent was part of the SNP QC procedures.)
 \item Pairwise Pearson $r^2\leq 0.1$ , to exclude SNPs in high linkage disequilibrium. (The $r^2$ coefficient was computed using \code{plink} and its \code{indep-pairwise} function with a moving window of size 1000 bp).
 \item Removed C/G and A/T SNPs to avoid unresolvable strand mismatches.
 \item Excluded SNPs in several regions with long-range LD \cite{price2008long}. (The list includes the MHC and 22 other regions.)
\end{itemize}

We computed the PCA loadings specifically on the individuals self-reported as British, to capture population structure only on this subset. The code used for this can be found at \code{GWAS\_pipeline/src/compute\_genomic\_PCs.R}.

\section*{GWAS on traditional cardiac indices}

GWAS was performed using the \code{bgenie} tool (version 1.3).

The volume of the different chambers was calculated as explained in \citep{ref_xia2022segm}. In particular, the volume for LV at end-diastole and end-systole was computed (LVEDV and LVESV).

The segmentation approach yields endocardial and epicardial surfaces for the LV, from which the myocardial mass of this chamber was calculated. This quantity is abbreviated LVM. Also, the left-ventricular mass-to-volume ratio (LVMVR) was computed as a the ratio between LVM and LVEDV.

We did not compute indexed versions of the previous phenotypes as does \cite{ref_pirruccello}, i.e. phenotypes divided by BSA, since we use this as covariate instead.

The sphericity index for the LV at end-diastole (LVEDSph) was estimated as follows: first, the convex hull (CH) of each cardiac mesh was obtained and, from it, its surface area $A_{\text{CH}}$ and volume $V_{\text{CH}}$. The sphericity was obtained as $\text{Sph}=(36\pi V_{\text{CH}})^{2/3}$, which is tantamount to the inverse of the ratio of $A_{\text{CH}}$ and the surface area of a sphere with volume $V_{\text{CH}}$. Since a sphere is the solid with minimal surface area, this number lies between 0 and 1. To obtain the CH and the associated areas and volumes, the submodule \code{Spatial} from the \code{SciPy} Python library was used (version 1.9.3).

\begin{figure}
    \centering
    \includegraphics[width=\linewidth]{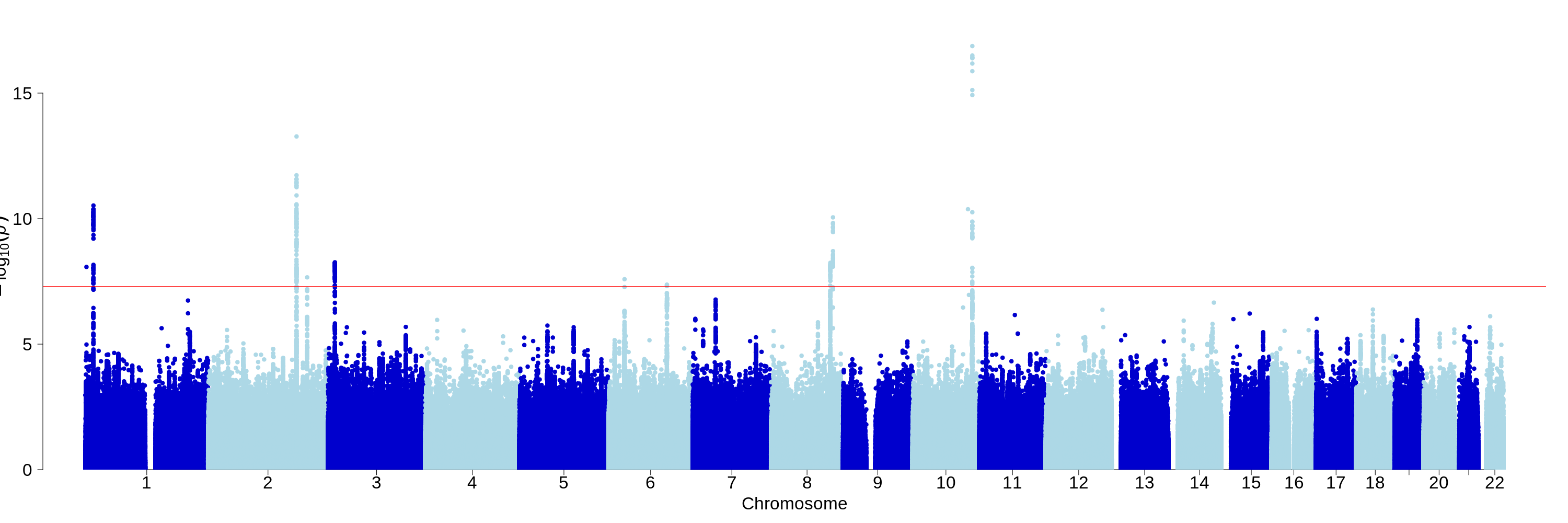}
    \caption{Manhattan plot of GWAS of left-ventricular end-systolic volume (LVESV).}
    \label{fig:LVESV}
\end{figure}

\begin{figure}
    \centering
    \includegraphics[width=\linewidth]{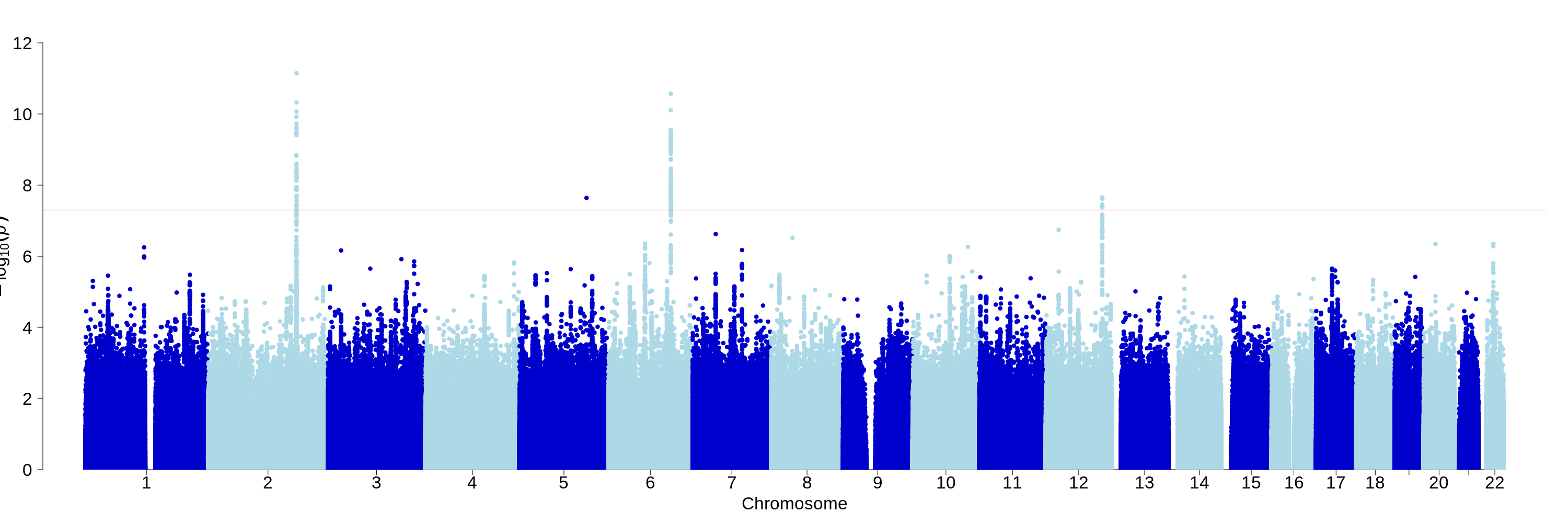}
    \caption{Manhattan plot of GWAS of left-ventricular myocardial mass (LVM).}
    \label{fig:LVM}
\end{figure}

\begin{figure}
    \centering
    \includegraphics[width=\linewidth]{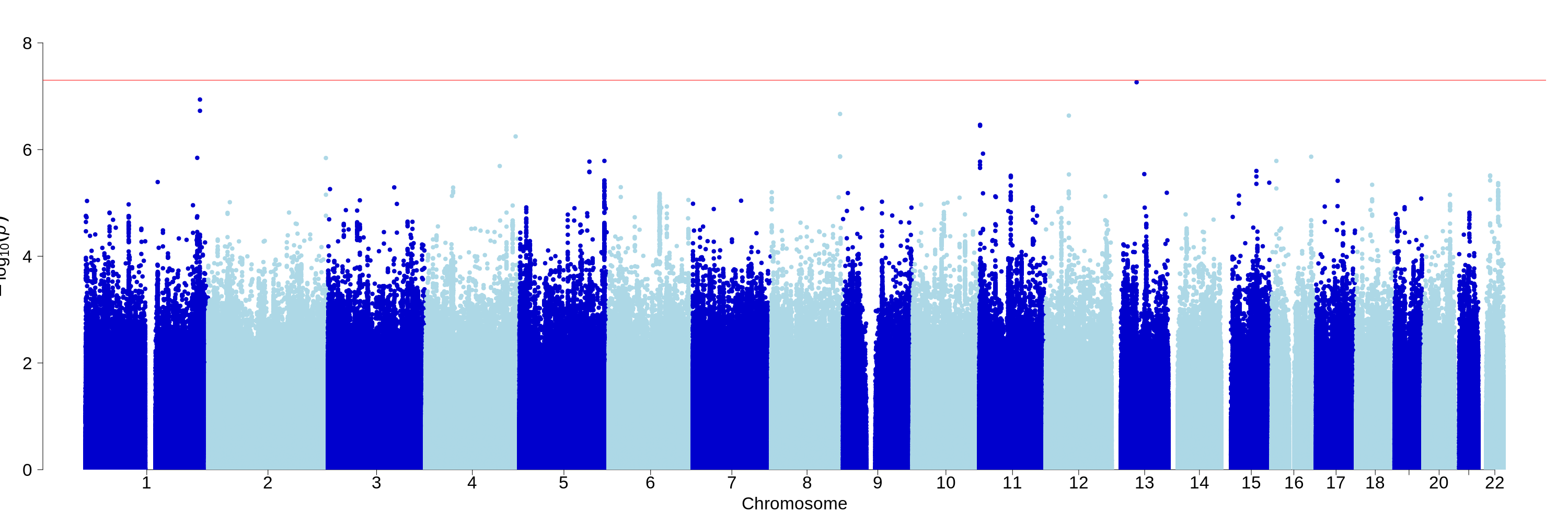}
    \caption{Manhattan plot of GWAS of left-ventricular mass-to-volume ratio (LVMVR).}
    \label{fig:LVMVR}
\end{figure}

GWAS was performed against these phenotypes, using the same covariates as with the unsupervised phenotypes. The corresponding Manhattan plots are displayed in Supplementary Figures \ref{fig:LVEDV}, \ref{fig:LVEDSph}, \ref{fig:LVESV}, \ref{fig:LVM} and \ref{fig:LVMVR}.

\begin{figure}
    \centering
    \includegraphics[width=\linewidth]{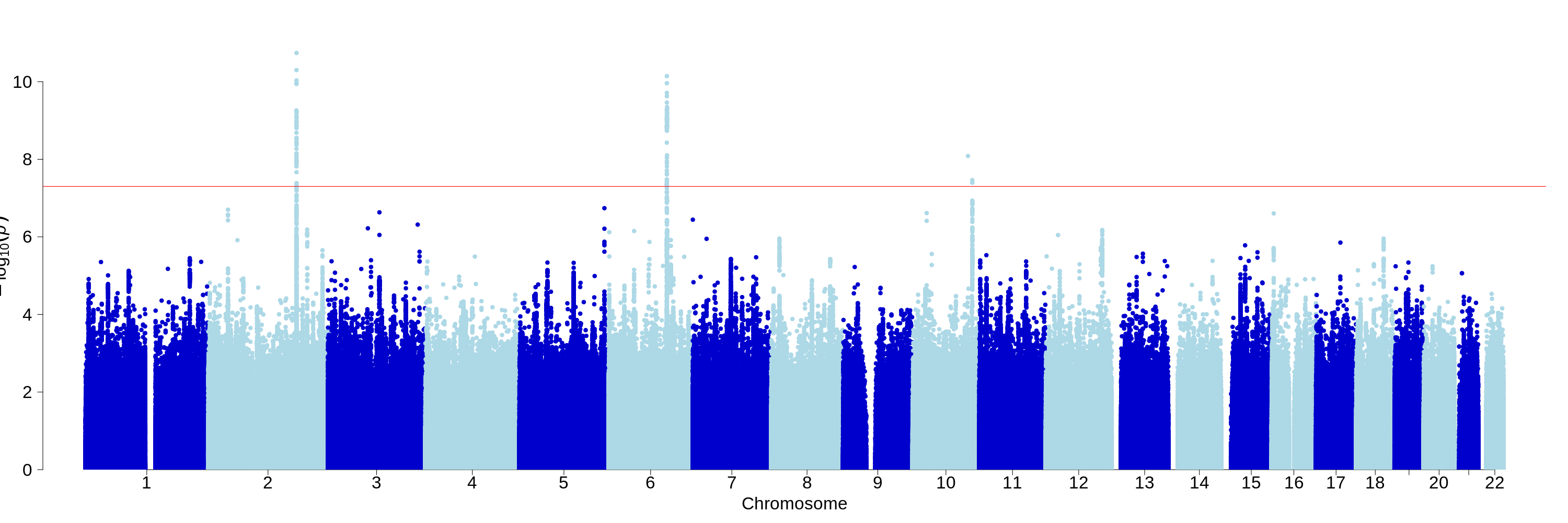}
    \caption{Manhattan plot of GWAS of left-ventricular end-diastolic volume (LVEDV)}
    \label{fig:LVEDV}
\end{figure}

\begin{figure}
    \centering
    \includegraphics[width=\linewidth]{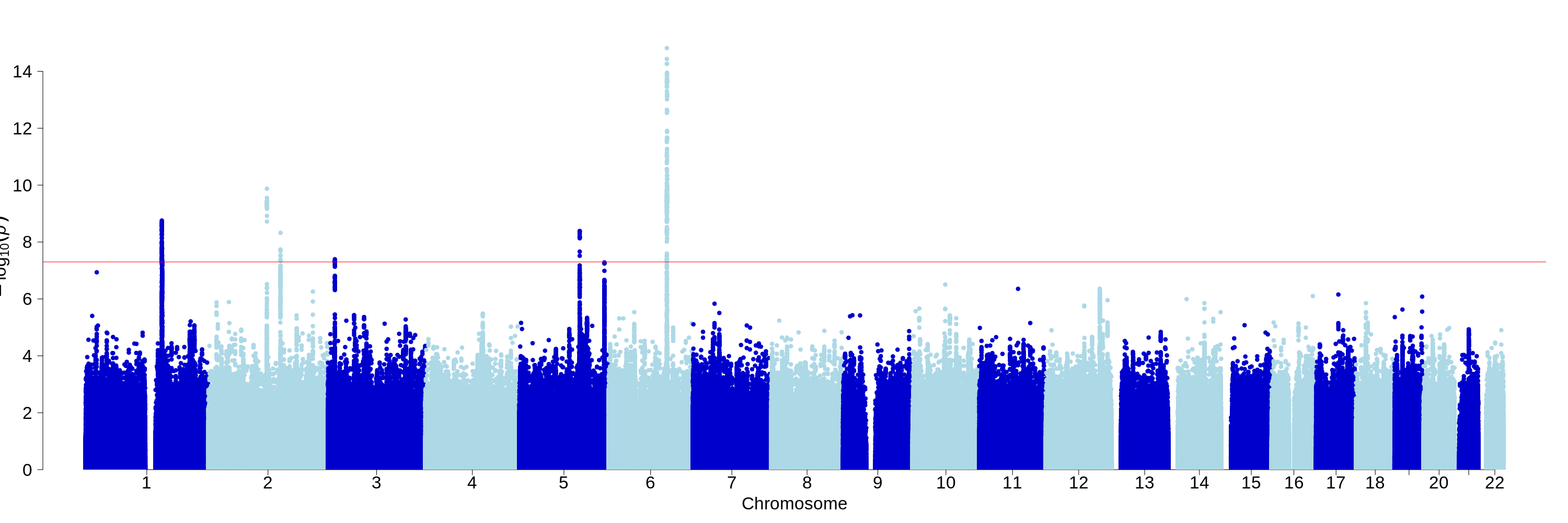}
    \caption{Manhattan plot of GWAS of left-ventricular end-diastolic sphericity (LVEDSph).}
    \label{fig:LVEDSph}
\end{figure}

\section*{Dimensionality reduction and genome-wide association studies (GWAS)}

Different dimensions of the latent space $n_z$ and weights $w_{\textrm{KL}}$ were studied, with the aim of achieving a compromise between reconstruction error and interpretability of the components.

\begin{figure}[ht!]
\includegraphics[width=\linewidth]{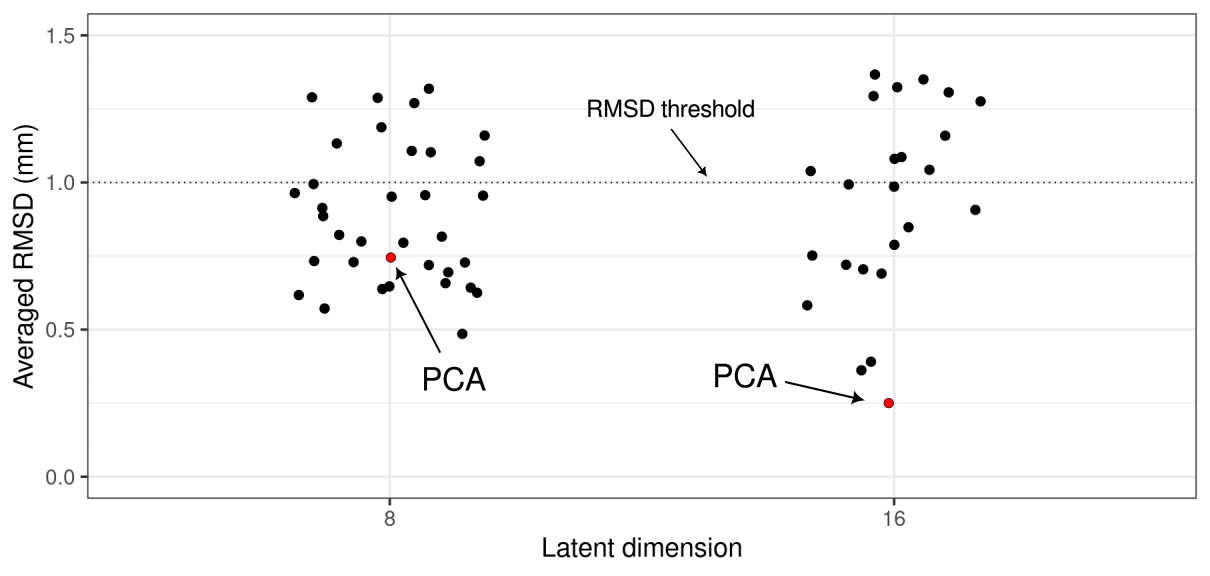}
\caption{Averaged reconstruction errors (measured as the RMSD averaged across the test set of 1000 subjects) for different CoMA models (black dots) and for PCA (red dots). A threshold of 1 mm in this metric is used to select the runs.}
\label{fig:reconstruction_performance}
\end{figure}

Figure \ref{fig:reconstruction_performance} shows a comparison of the reconstruction error obtained through principal component analysis (PCA) and convolutional mesh autoencoders (CoMA), as a function of the number of the components $n_z$ of the latent space (values of 8 and 16 were used). CoMA and PCA yielded comparable reconstruction errors, with the best CoMA  runs outperforming PCA slightly for $n_z=8$, and PCA outperforming CoMA for $n_z=16$. No significant difference was found between non-variational and variational CoMA in terms of the reconstruction error. The distribution of RMSD per subject can be examined for each run through the Shiny app.

\begin{figure}
    \centering
    \includegraphics[width=0.7\linewidth]{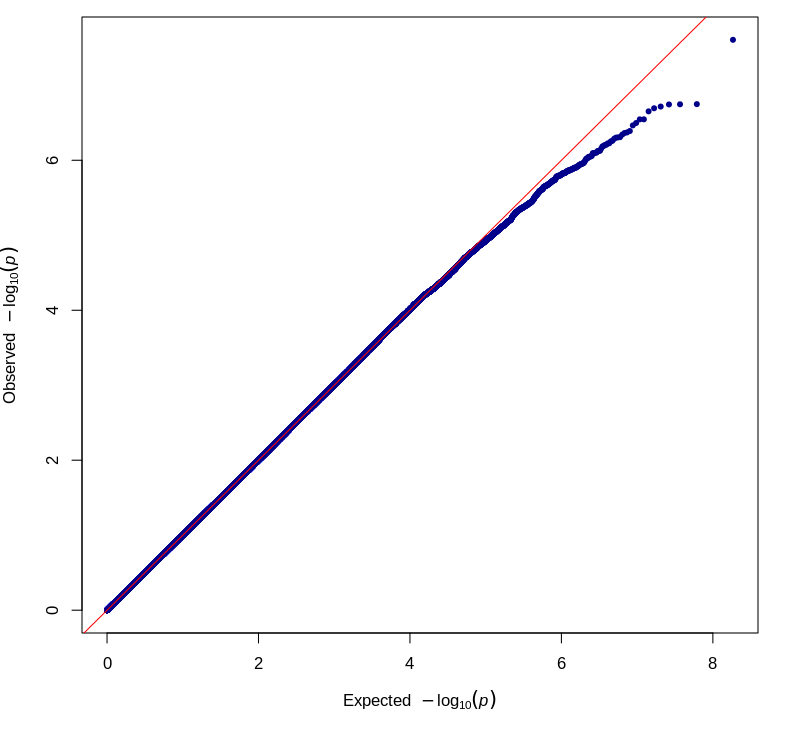}
    \caption{Q-Q plot for the GWAS on the 16 first shape PCs (the associations for each PC have been pooled into a single plot). We see that, despite the good reconstruction performance, PCA fails to produce a representation with a genetic component.}
    \label{fig:pca_gwas}
\end{figure}

\subsection*{Convolutional mesh autoencoder: implementation details}
For the autoencoder, we ran a grid optimisation scheme for meta-parameter selection, both for the network architecture and the training process. For each number of components $n_z$ and regularisation weight $w_\textrm{KL}$, the execution that presented the minimum mean squared error (MSE) within the validation set was chosen. The autoencoder architecture is detailed in Supplementary Table S2. 
After each convolutional layer, a ReLU activation function was applied. The number of samples used for training was 5,000, whereas the validation set contained 1,000 individuals. The sample partition, as well as the weight initialisation and the sampling process of the VAE, was controlled by a random seed that was stored along with the trained model for reproducibility; 3 to 5 different random seeds were utilised for each parameter configuration. The Adam optimiser was used to find optimal network parameters, by minimising the KL-regularised MSE reconstruction loss \cite{kingma2014adam}. The learning rate that achieved good performance utilised were in the range $[10^{-4}, 3\times10^{-4}]$.

The network training was performed on a Nvidia DGX A100 workstation located at the University of Leeds. This machine is endowed with Nvidia A100 GPUs.
The genetic pipeline was executed on the University of Leeds' high performance computing cluster, ARC. The workload has been parallelised across computing nodes using the Son of the Grid Engine (SGE) queue management system.

\begin{table}
\begin{center}
\begin{tabular}{|c|c|c|}
\hline
          & \textbf{Input} & \textbf{Output} \\ \hline
ChebConv  & $5220\times 3$ &  $5220\times C_1$ \\ \hline
DS        & $5220\times C_1$ & $2610\times C_1$ \\ \hline
ChebConv  & $2610\times C_1$ & $2610\times C_2$ \\ \hline
DS        & $2610\times C_2$ & $1305\times C_2$ \\ \hline
ChebConv  & $1305\times C_2$ & $1305\times C_3$ \\ \hline
DS        & $1305\times C_3$ & $652\times C_3$\\ \hline
ChebConv  & $652\times C_3$ &  $652\times C_4$\\ \hline
DS        & $652\times C_4$ &  $326\times C_4$\\ \hline
ChebConv  & $326\times C_4$ &  $326\times C_5$\\ \hline
FC        & $326\times C_5$ &  $n_z\times 1$\\ \hline
\end{tabular}
\end{center}
\caption{Architecture of the encoder part used for each of the cardiac chambers. The decoder has the same architecture but reading from the bottom upwards and inverting input and output. (ChebConv: Chebyshev convolution, DS: downsampling, FC: fully connected layer.) The architectural hyperparameters were $(C_1, C_2, C_3, C_4, C_5)\in\{(16, 32, 64, 128), (128, 128, 128, 128), (1024, 512, 256, 128)\}$ and $n_z\in\{8, 16\}$.}
\label{table:AE_arch}
\end{table}

\subsection*{Locus-level figures.} 
Here we present, for each locus, a triad of plots which consists of: 1) the Manhattan plot for the GWAS of the single phenotype (i.e. latent variable) that yielded the strongest $p$-value against that locus, 2) a LocusZoom plot of a 1Mb region centered at the lead SNP for each locus and 3) a sequence of meshes for different ranges of quantiles of the latent variable. Note that all of these plots can be also queried using the Shiny app (see \url{www.github.com/rbonazzola/LV-GWAS-ShinyApp}).

\paragraph{Manhattan plots.} Manhattan plots were generated using the R package called \code{qqman}. Code for this can be found at \code{GWAS\_pipeline/src/postprocessing/gwas\_analysis.R}.

\paragraph{LocusZoom plot.}
LocusZoom plots were generated using the 1.4 version of the tool (which is not currently maintained anymore). These plots allow to examine the region close to the lead SNPs, see which genes lie in that region; and also, by examining the correlation with nearby SNPs, to visually determine the presence of independent GWAS signals. The LD reference utilised was the European subpopulation of the 1000 Genomes panel (March 2012). Note that some of the genetic variants in the UKB SNP microarray had no available LD information; in that case, they are coloured in grey. Moreover, the plots are annotated with GWAS hits from previous GWAS summary statistics hosted at \code{www.genome.gov}; however, bear in mind that this list is far from complete. A Python script was used to generate the LocusZoom plot commands for each of the loci and latent variables of interest and it can be found at: \code{CardiacGWAS/analysis/locuszoom.py}.

\paragraph{Morphological interpretation.} An interpretation of the impact on LV morphology of the latent variables linked to different loci was achieved by examining the average shape of subjects located at different quantiles. Prior to averaging, the sets of meshes were unscaled, and then scaled back after averaging. The quantile ranges used were: [0, 0.01], [0.095, 0.105], [0.495, 0.505], [0.895, 0.905] and [0.99, 1]. Note that, since there are \NCMRGBR  subjects in our database, each quantile range (of 1\% in width) encompasses more than 300 subjects. In all cases, we observe a smooth transition in shape from lower to higher values. 

An alternative approach that was tested was the following: varying the components of the latent representation one at a time (while keeping the others fixed at the mean value) and generating the associated synthetic shapes by means of the trained decoder. However, note that since $w_{\text{KL}}$ (the parameter that controls the strength of the variational loss) spans a broad range of values in our ensemble of runs, independence of the different latent variables within a run is not guaranteed for all runs (and indeed, it is not observed in most of them). For this reason, it is not possible in general to use the decoder in this way: it is only valid when statistical independence of the latent variables is observed.

Additionally, the Spearman correlation coefficients between these latent variables and the four LV handcrafted phenotypes are provided in Table S3.

\begin{table*}[ht!]
\begin{center}
\begin{tabular}{lrrrr}
\toprule
{} &  LVEDV &  LVEDSph &    LVM &  LVMVR \\
\midrule
PLN    &  0.434 &    0.625 &  0.333 & -0.158 \\
GOSR2  & -0.335 &   -0.272 & -0.397 & -0.148 \\
TTN    &  0.889 &   -0.157 &  0.904 &  0.224 \\
TBX5   &  0.888 &    0.000 &  0.761 & -0.035 \\
NKX2.5 & -0.865 &   -0.372 & -0.790 &  0.002 \\
LMO7   &  0.927 &   -0.014 &  0.832 &  0.021 \\
RBM20  &  0.971 &    0.070 &  0.887 &  0.040 \\
CNOT7 & -0.900 &   -0.135 & -0.902 & -0.191 \\
WAC    &  0.888 &   -0.120 &  0.903 &  0.199 \\
LGALS8  &  0.672 &    0.227 &  0.605 & -0.061 \\
CCDC91 &  0.499 &    0.248 &  0.358 & -0.156 \\
EN1    & -0.041 &   -0.461 & -0.137 & -0.112 \\
BAG3   &  0.928 &    0.089 &  0.841 &  0.018 \\
OR9Q1  &  0.528 &    0.603 &  0.440 & -0.107 \\
STRN   & -0.165 &   -0.429 & -0.183 & -0.296 \\
\bottomrule
\end{tabular}
\caption{Spearman correlation between the best phenotypes per locus and four LV handcrafted indices.}
\end{center}
\label{table:corr_with_indices}
\end{table*}

\begin{figure*}[ht!]
\centering
\includegraphics[width=\textwidth]{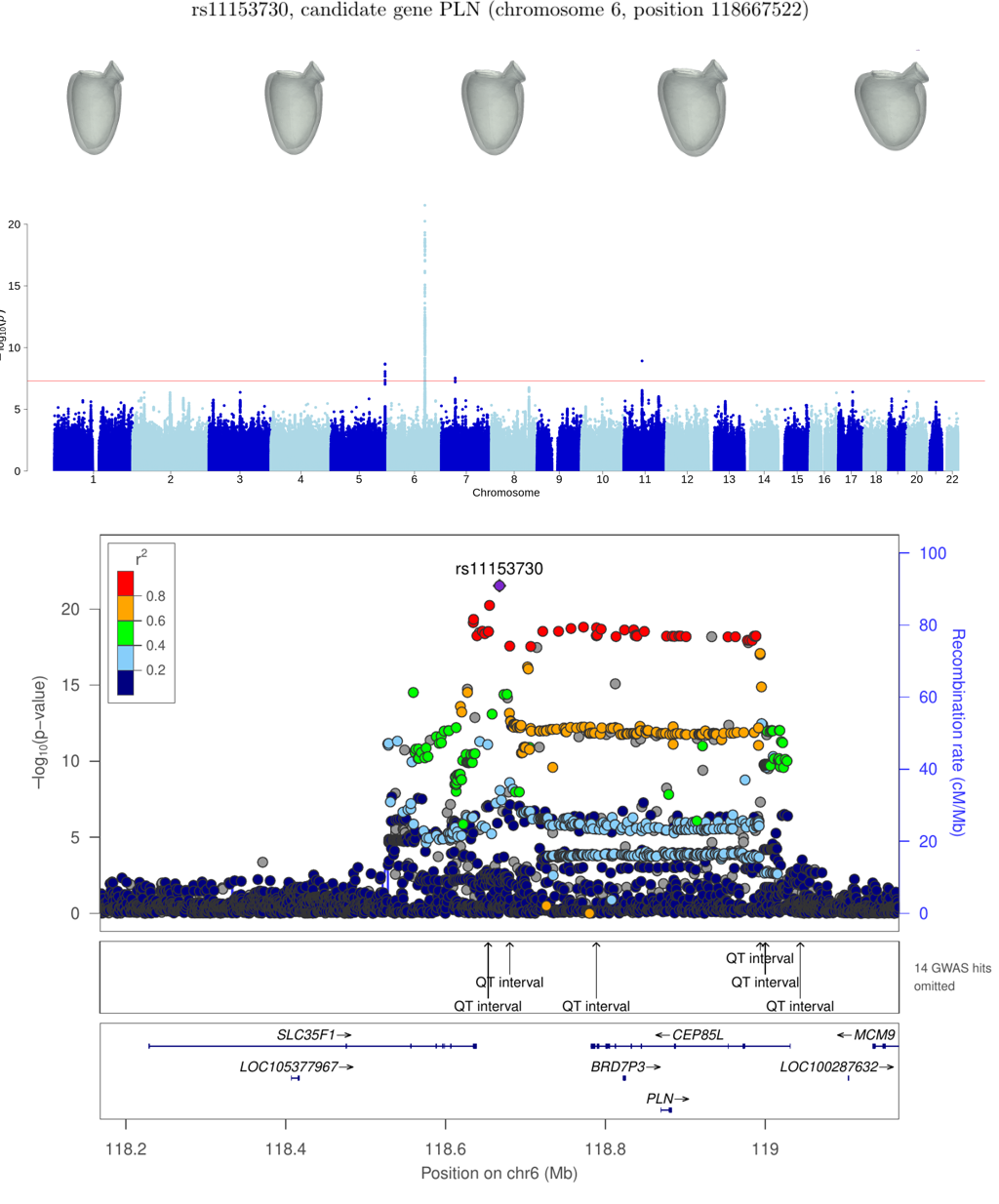}
\caption{Triad of plots for locus PLN.}
\label{fig:chr6_78}
\end{figure*}

\begin{figure*}[ht!]
\centering
\includegraphics[width=\textwidth]{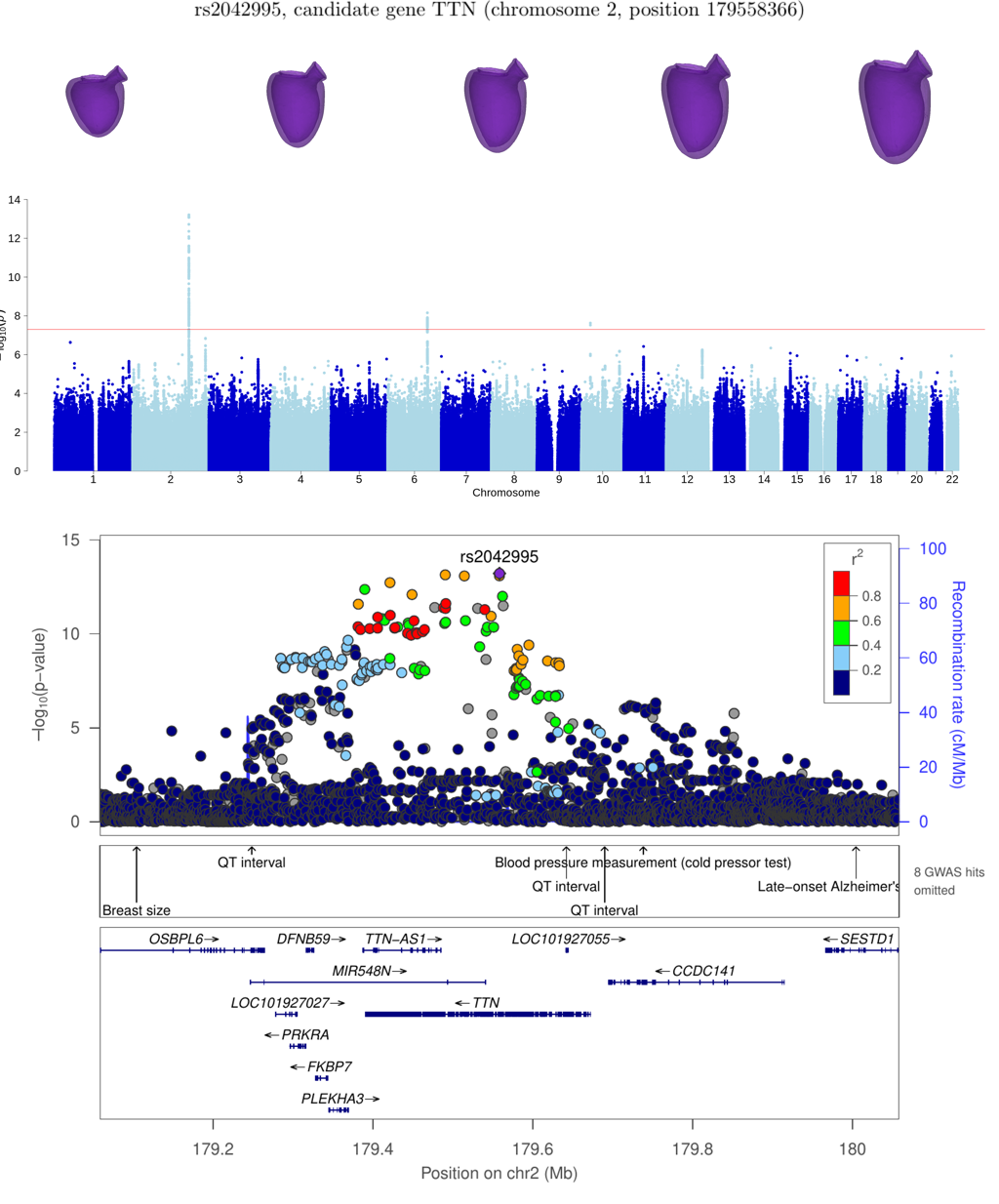}
\caption{Triad of plots for locus TTN.}
\label{fig:chr2_108}
\end{figure*}

\begin{figure*}[ht!]
\centering
\includegraphics[width=\textwidth]{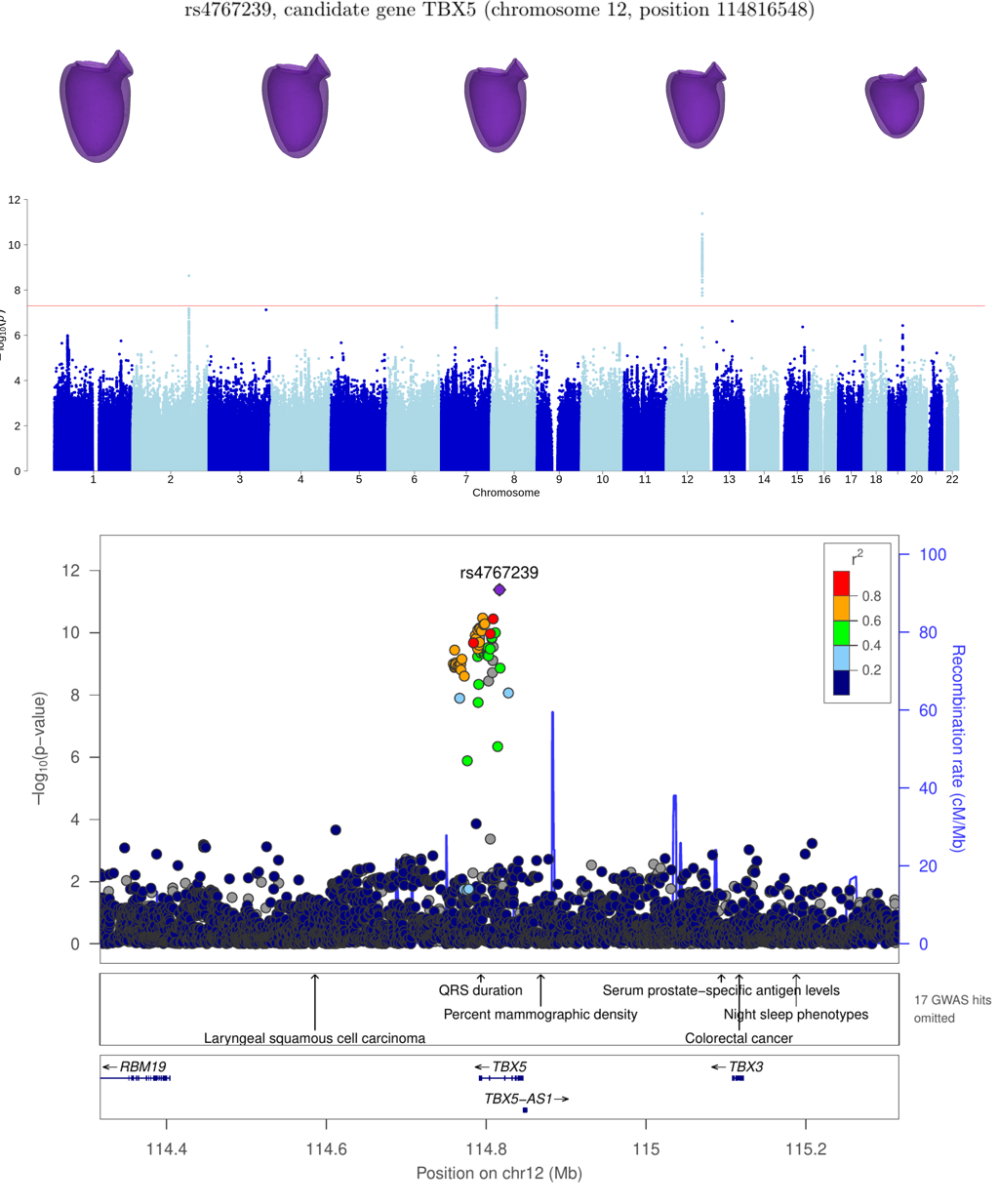}
\caption{Triad of plots for locus TBX5.}
\label{fig:chr12_69}
\end{figure*}

\begin{figure*}[ht!]
\centering
\includegraphics[width=\textwidth]{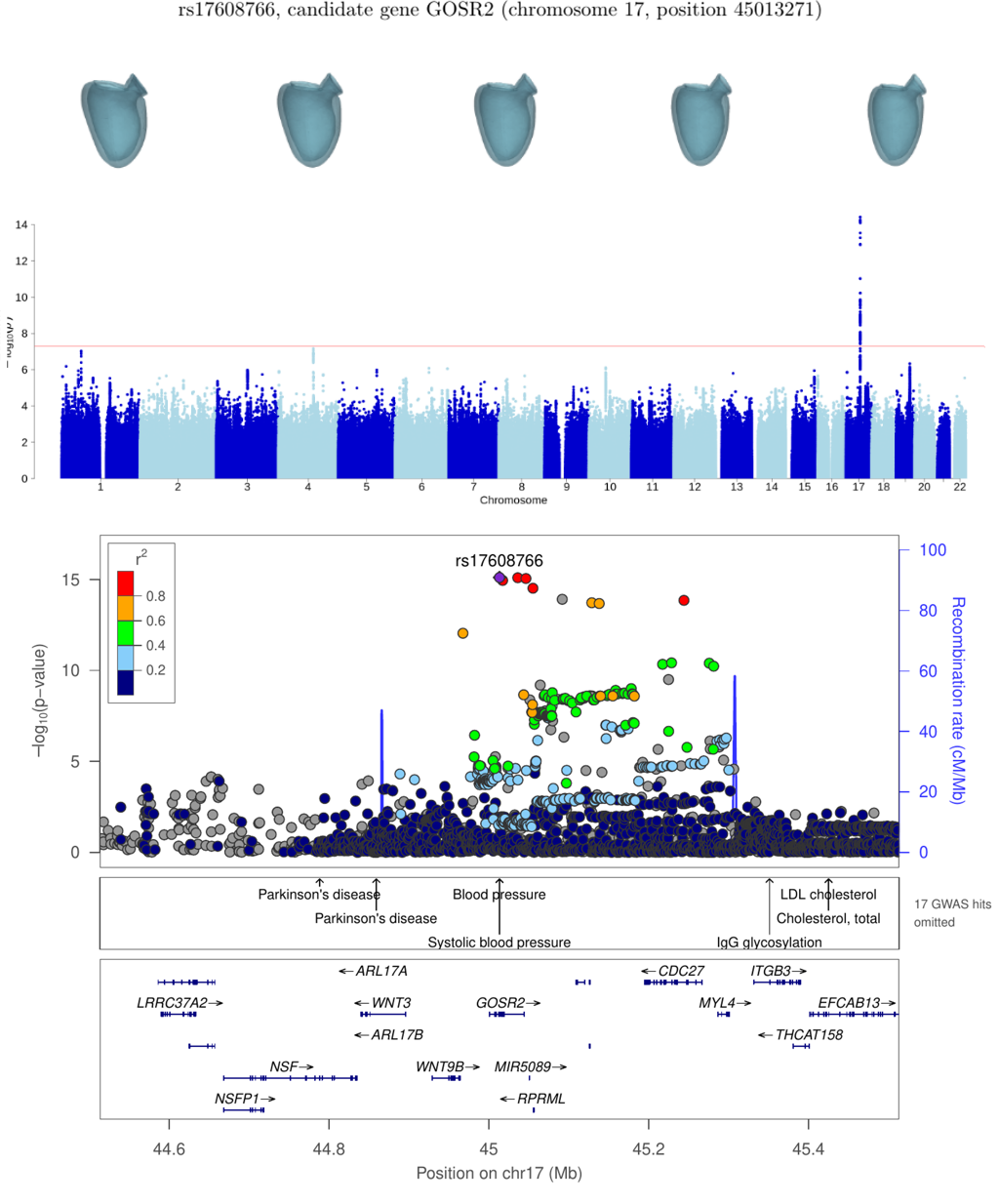}
\caption{Triad of plots for locus GOSR2.}
\label{fig:chr17_27}
\end{figure*}

\begin{figure*}[ht!]
\centering
\includegraphics[width=\textwidth]{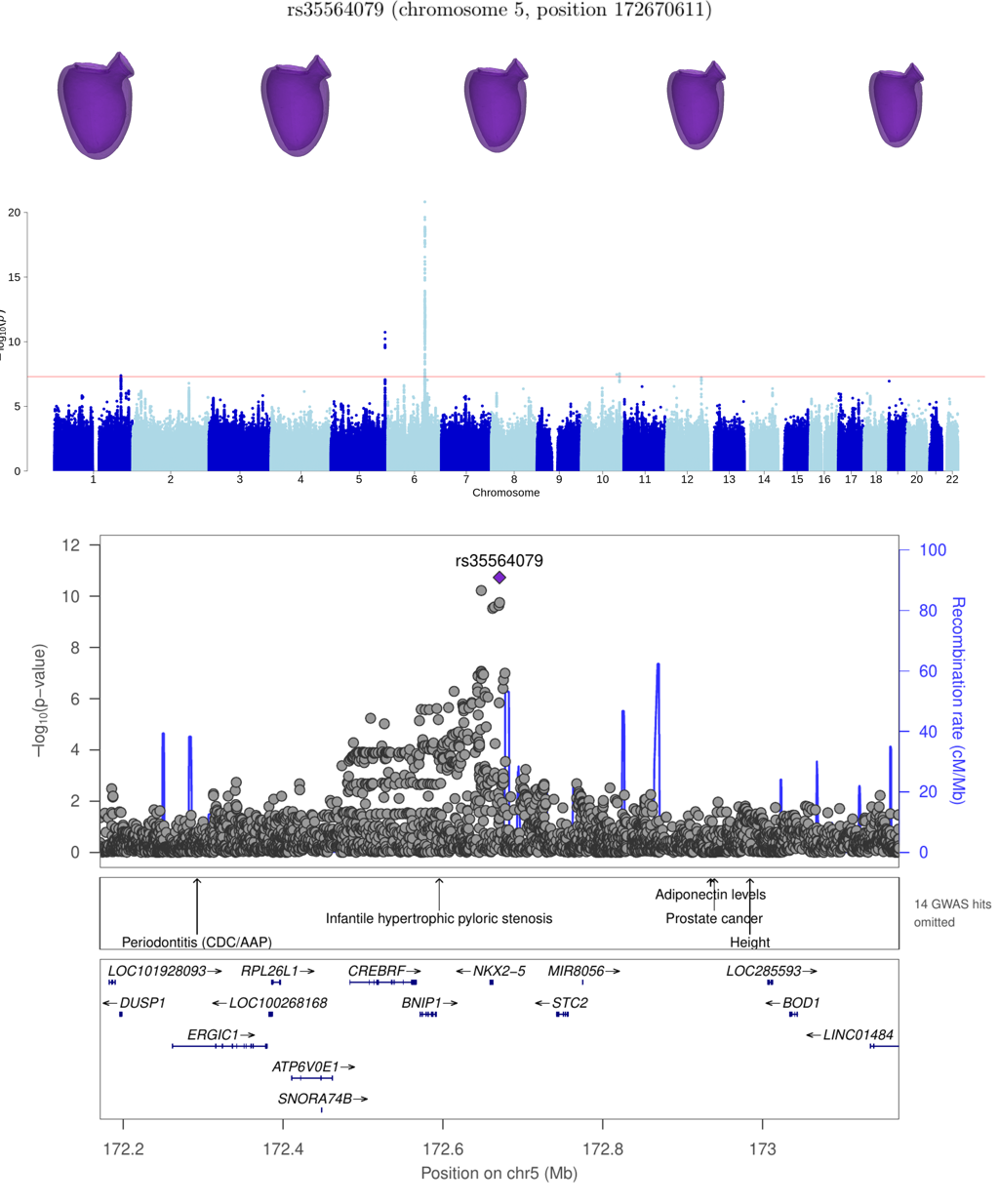}
\caption{Triad of plots for locus NKX2.5.}
\label{fig:chr5_103}
\end{figure*}

\begin{figure*}[ht!]
\centering
\includegraphics[width=\textwidth]{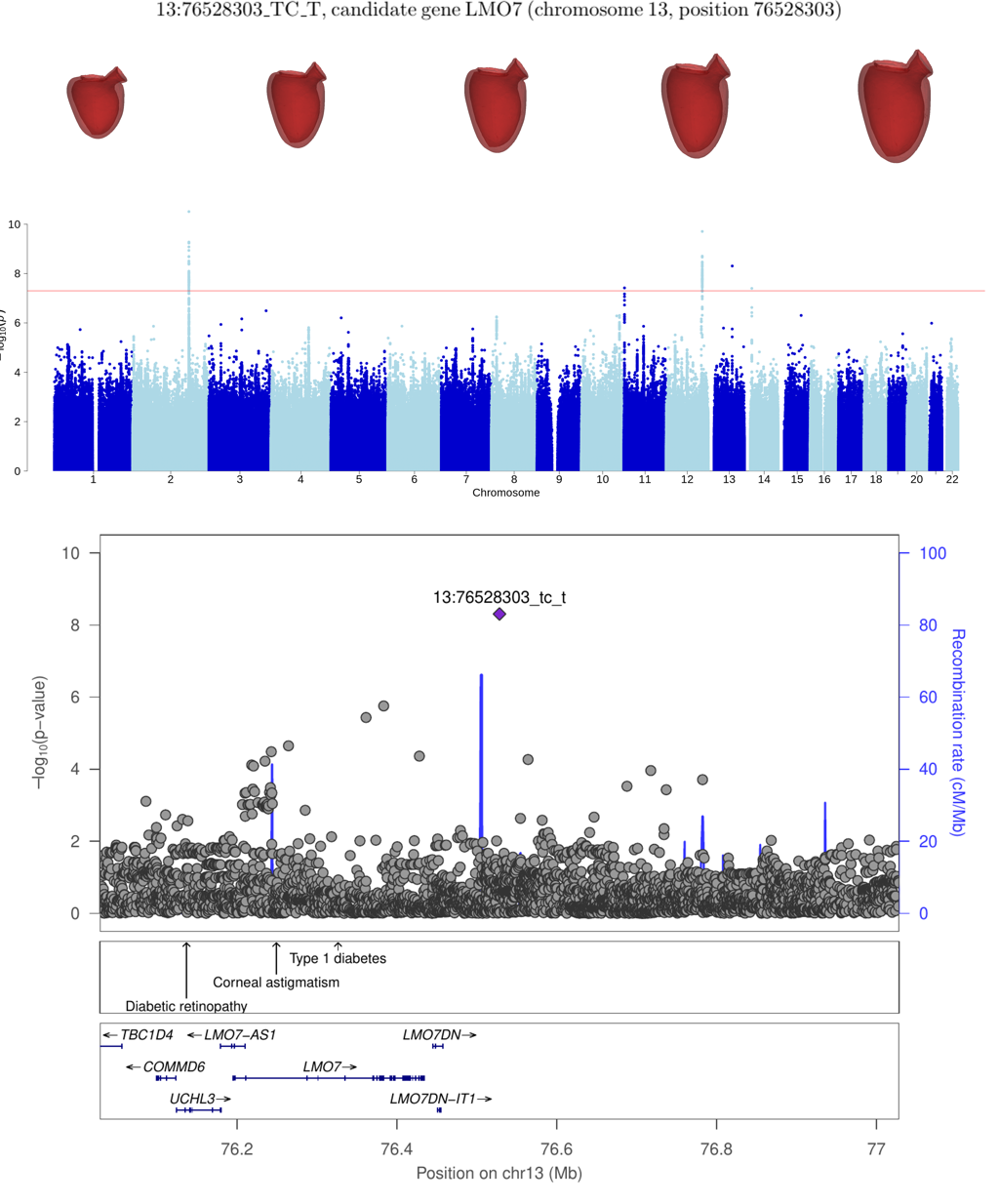}
\caption{Triad of plots for locus LMO7.}
\label{fig:chr13_37}
\end{figure*}

\begin{figure*}[ht!]
\centering
\includegraphics[width=\textwidth]{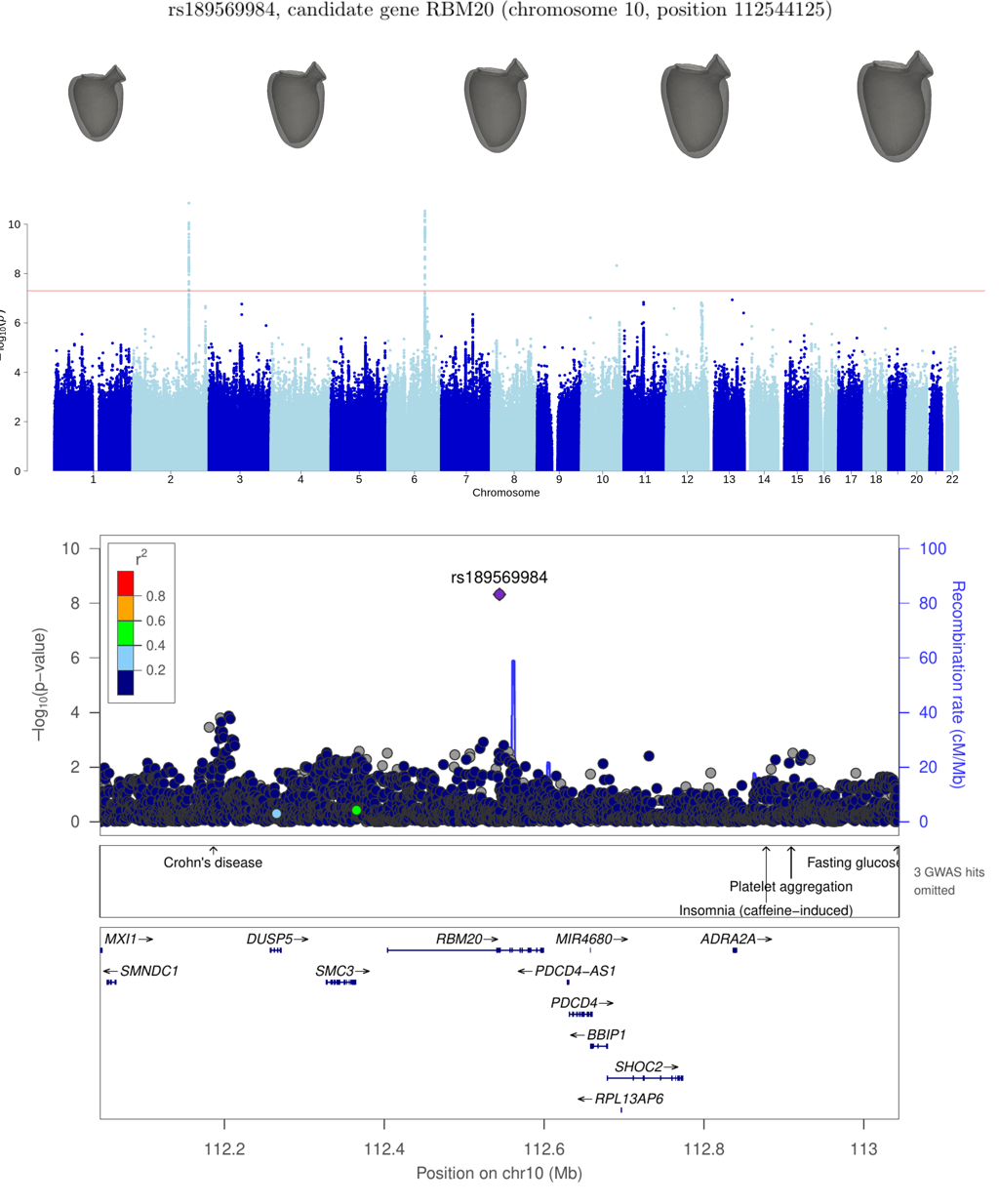}
\caption{Triad of plots for locus RBM20.}
\label{fig:chr10_69}
\end{figure*}

\begin{figure*}[ht!]
\centering
\includegraphics[width=\textwidth]{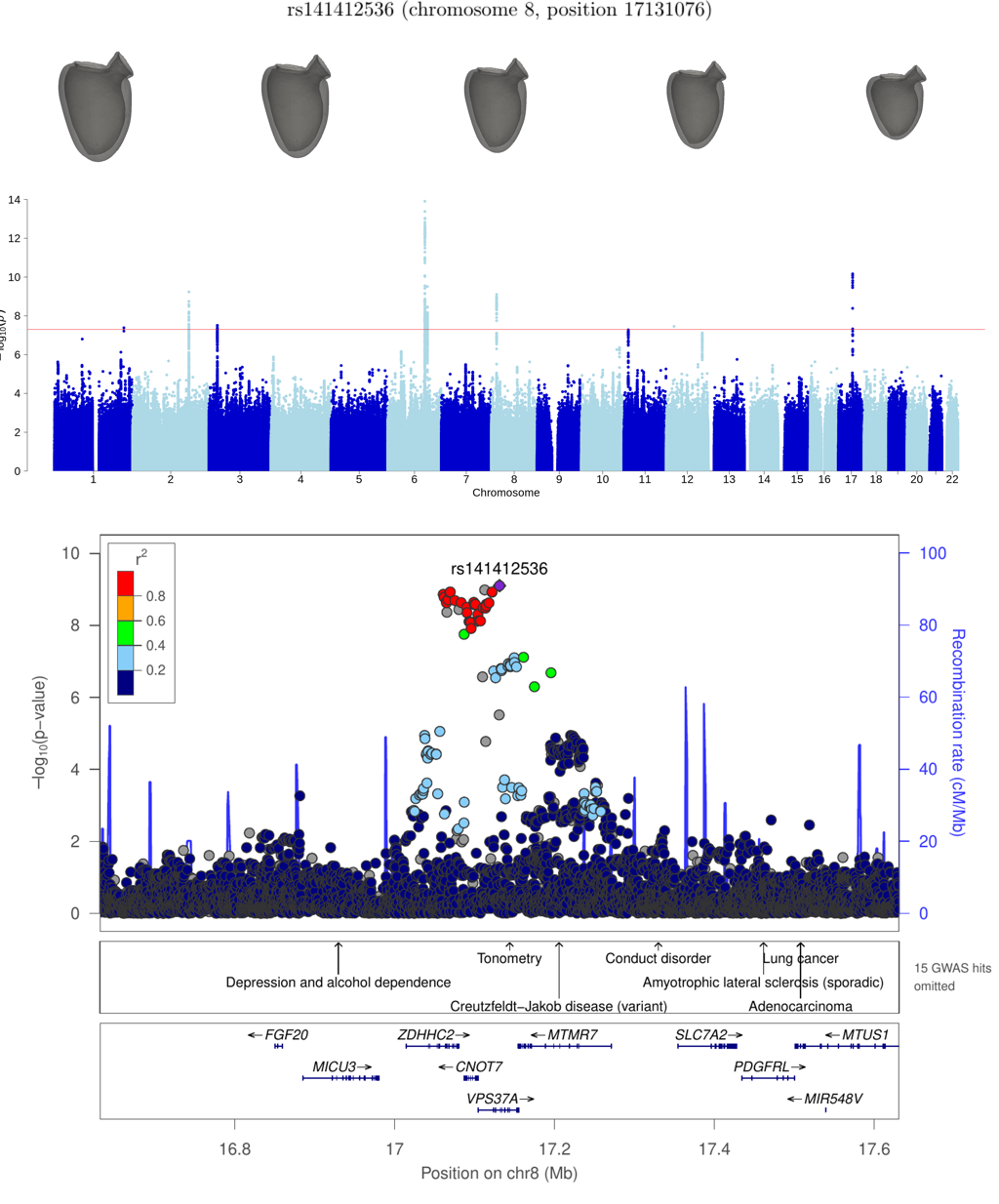}
\caption{Triad of plots for locus CNOT7.}
\label{fig:chr8_18}
\end{figure*}

\begin{figure*}[ht!]
\centering
\includegraphics[width=\textwidth]{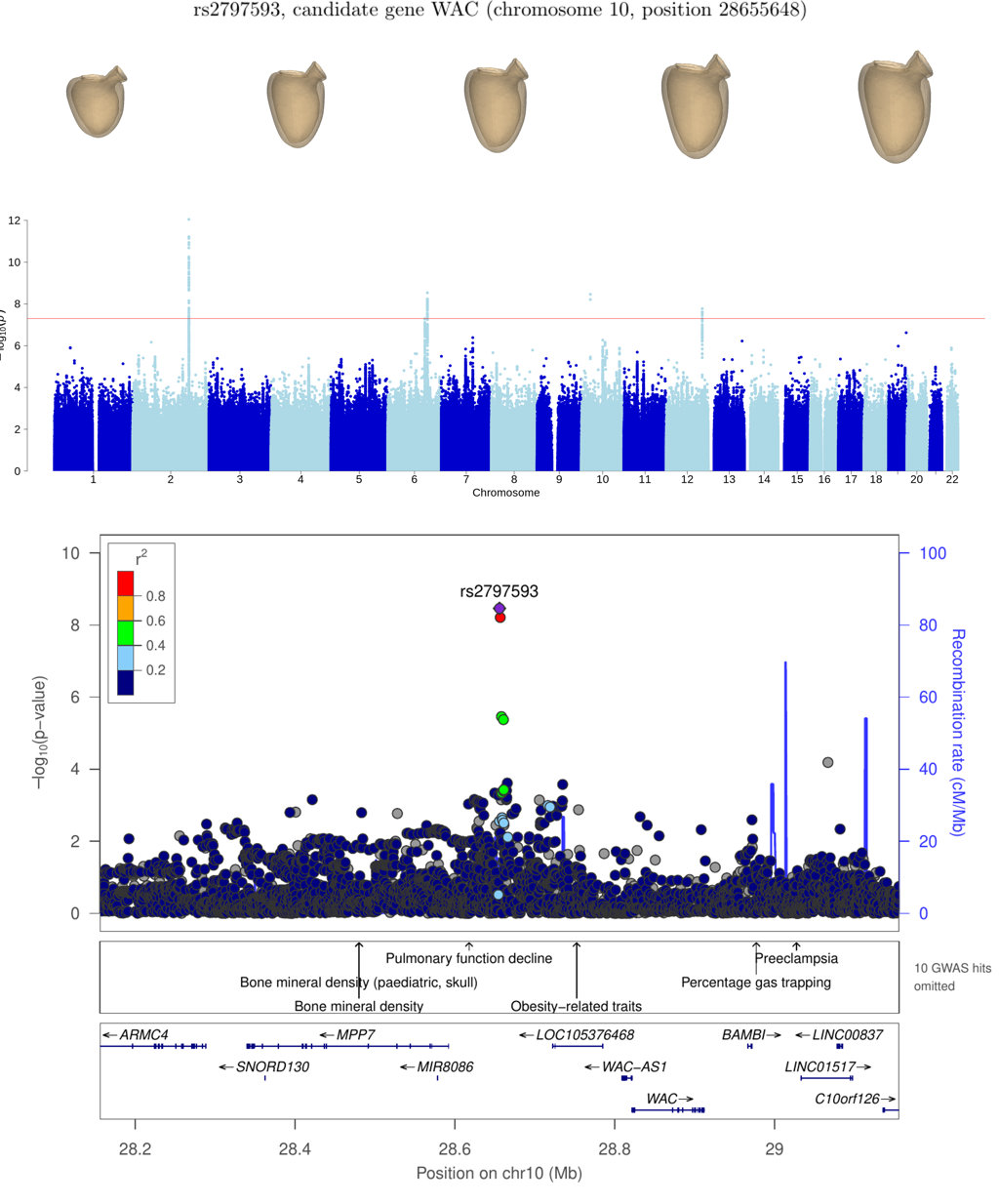}
\caption{Triad of plots for locus WAC.}
\label{fig:chr10_20}
\end{figure*}

\begin{figure*}[ht!]
\centering
\includegraphics[width=\textwidth]{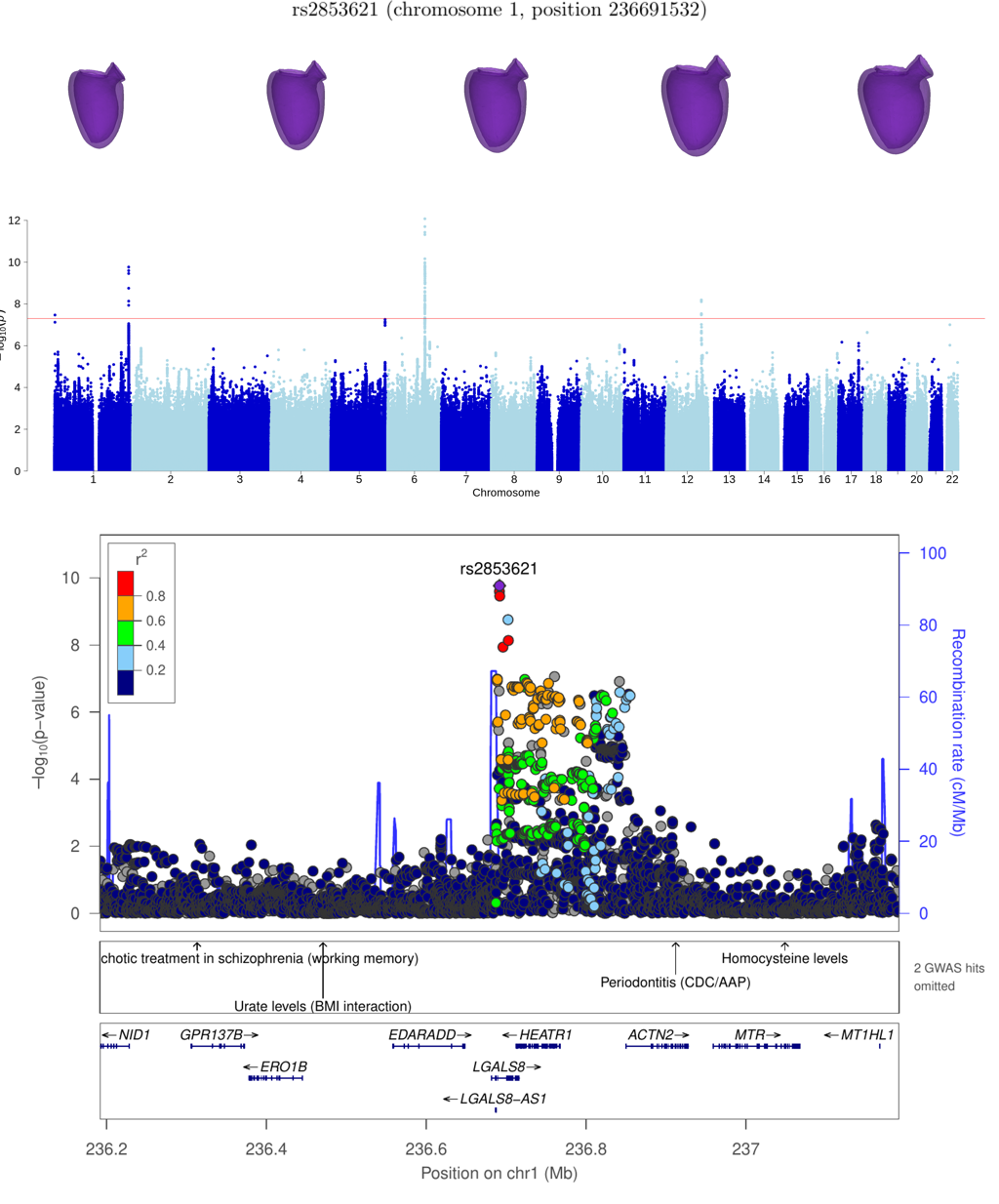}
\caption{Triad of plots for locus near gene LGALS8.}
\label{fig:chr1_124}
\end{figure*}

\begin{figure*}[ht!]
\centering
\includegraphics[width=\textwidth]{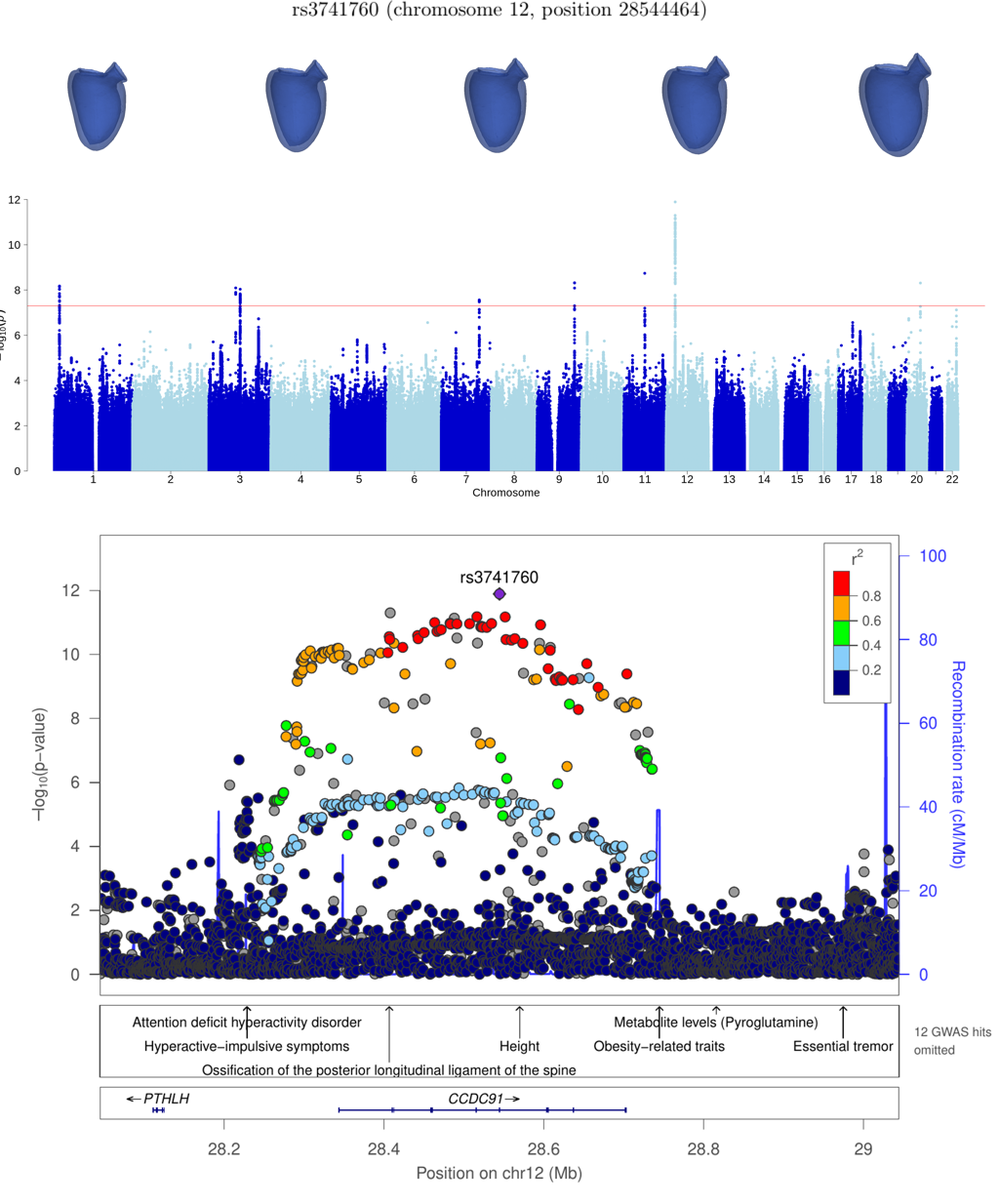}
\caption{Triad of plots for locus near gene CCDC91.}
\label{fig:chr12_19}
\end{figure*}

\begin{figure*}[ht!]
\centering
\includegraphics[width=\textwidth]{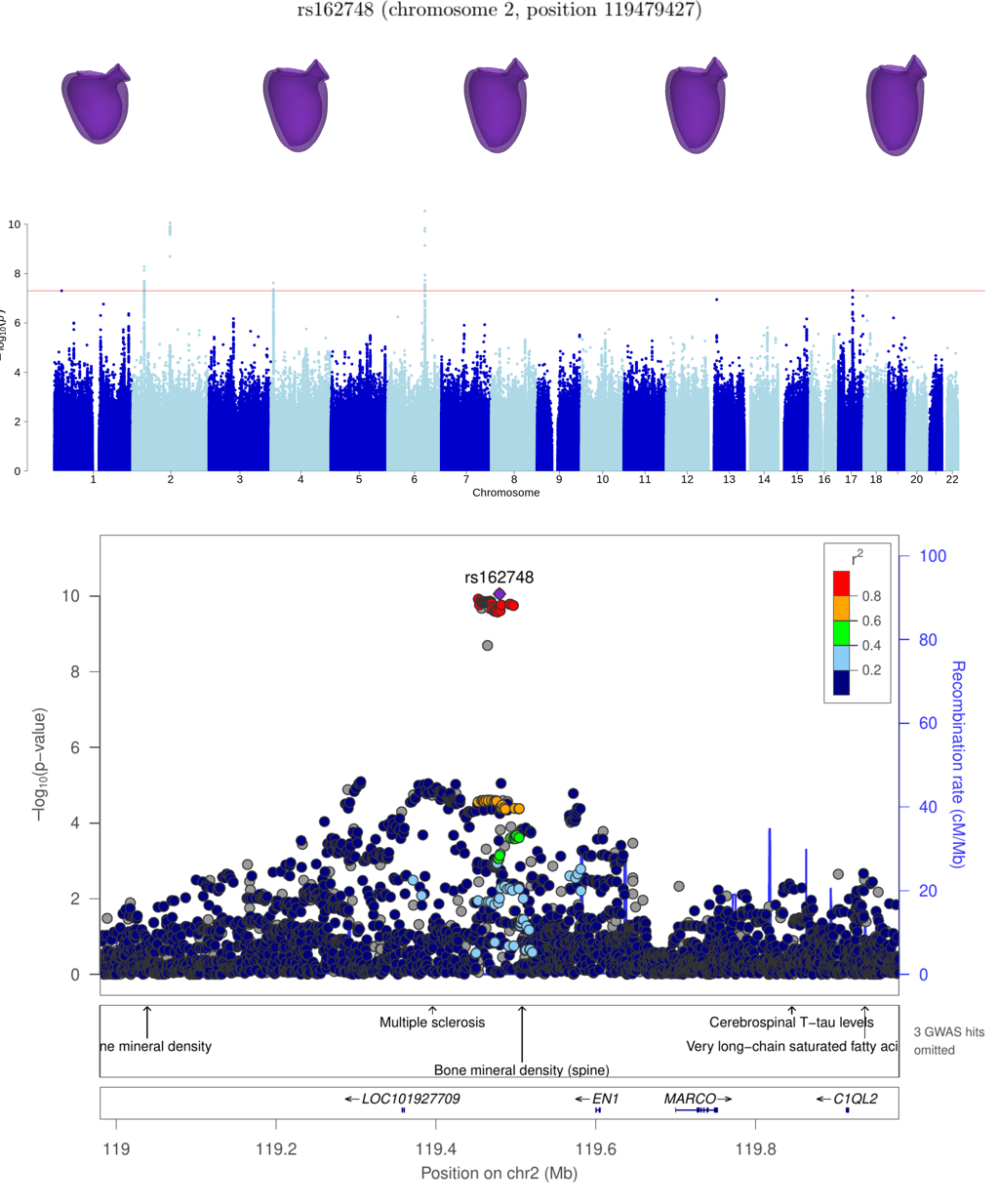}
\caption{Triad of plots for locus near gene EN1.}
\label{fig:chr2_69}
\end{figure*}

\begin{figure*}[ht!]
\centering
\includegraphics[width=\textwidth]{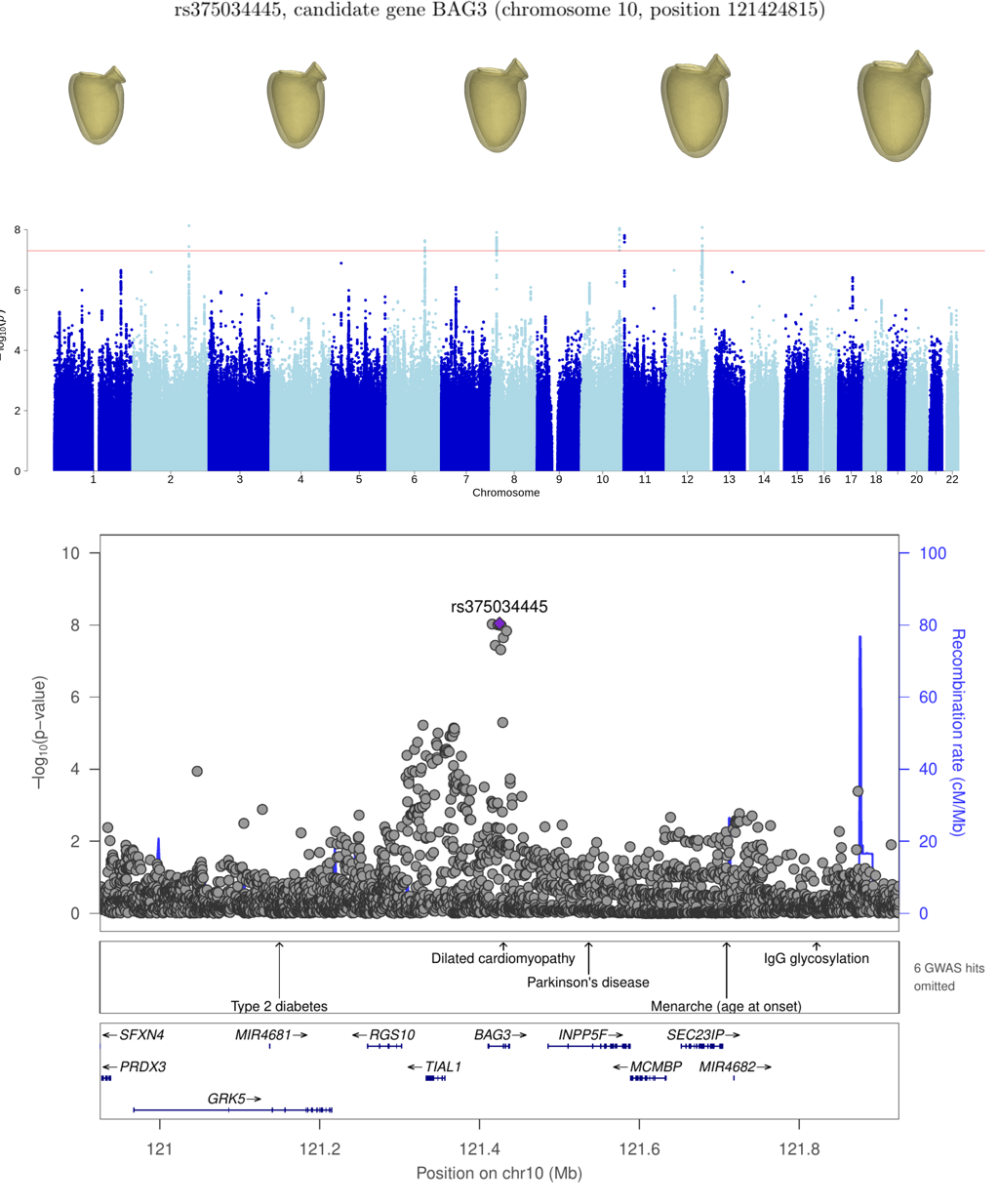}
\caption{Triad of plots for locus near gene BAG3.}
\label{fig:chr10_74}
\end{figure*}

\begin{figure*}[ht!]
\centering
\includegraphics[width=\textwidth]{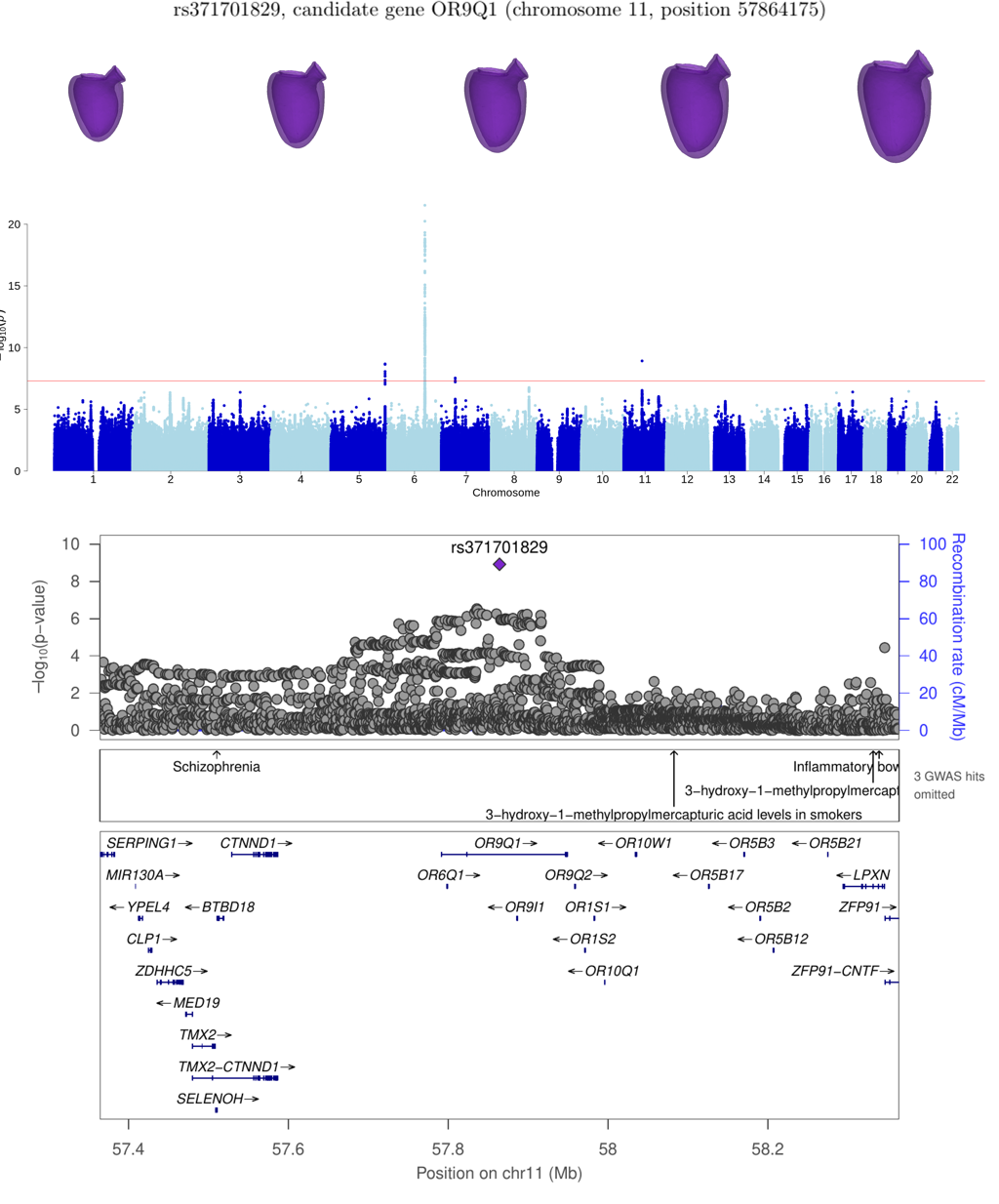}
\caption{Triad of plots for locus near gene OR9Q1.}
\label{fig:chr11_32}
\end{figure*}

\begin{figure*}[ht!]
\centering
\includegraphics[width=\textwidth]{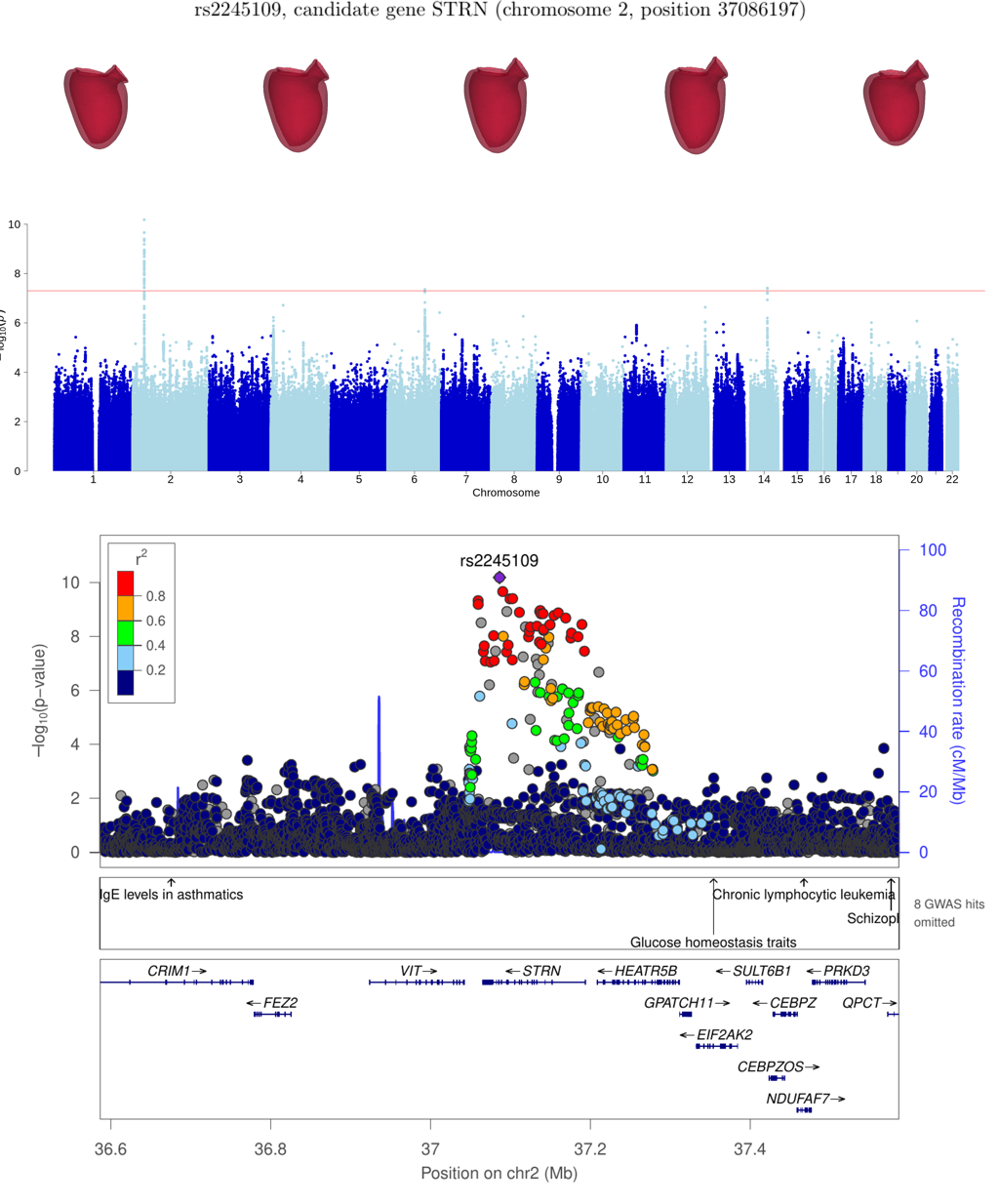}
\caption{Triad of plots for locus STRN.}
\label{fig:chr2_23}
\end{figure*}

\end{document}